%% file: main.tex
\documentclass[letterpaper,twocolumn,10pt]{article}
\usepackage{usenix,epsfig,endnotes}
\usepackage{multirow}
\newcommand{\paragraphX}[1]{\vskip 4pt \noindent \textbf{#1} \hskip .05in}

\usepackage{balance}
\usepackage[ruled]{algorithm2e} % For algorithms

\usepackage{amsmath}
\usepackage{easy-todo}
\usepackage{graphicx}
\usepackage{caption}
\usepackage[english]{babel} % handle hyphenation
\usepackage[caption=false,font=footnotesize]{subfig}

\newcommand{\squishlist}{
 \begin{list}{\textendash}
  { \setlength{\itemsep}{0pt}
     \setlength{\parsep}{3pt}
     \setlength{\topsep}{3pt}
     \setlength{\partopsep}{0pt}
     \setlength{\leftmargin}{1em}
     \setlength{\labelwidth}{0.7em}
     \setlength{\labelsep}{0.4em} } }

\newcommand{\squishend}{
  \end{list}  }
\begin{document}

\newenvironment{smenum}{
\begin{enumerate}
  \setlength{\itemsep}{1pt}
  \setlength{\parskip}{0pt}
  \setlength{\parsep}{0pt}
  \setlength{\topsep}{3pt}
}{\end{enumerate}}

%don't want date printed
%\date{}

%make title bold and 14 pt font (Latex default is non-bold, 16 pt)

\title{\Large \bf Towards Predicting Efficient and Anonymous \\ Tor Circuits}

%for single author (just remove % characters)
\author{
{\rm Armon Barton}\\
University of Texas at Arlington
\and
{\rm Mohsen Imani}\\
University of Texas at Arlington
 %copy the following lines to add more authors
 \and
 {\rm Jiang Ming}\\
University of Texas at Arlington
\and
 {\rm Matthew Wright}\\
Rochester Institute of Technology
} % end author

\maketitle

% Use the following at camera-ready time to suppress page numbers.
% Comment it out when you first submit the paper for review.
%\thispagestyle{empty}

\subsection*{Abstract}
\input{MyFiles/abstract}

\input{MyFiles/Intro}

\input{MyFiles/background.tex}

\input{MyFiles/classification.tex}
\input{MyFiles/predictor.tex}
\input{MyFiles/liveTor.tex}

\input{MyFiles/anonymity.tex}
\input{MyFiles/security.tex}

\input{MyFiles/disc.tex}

\input{MyFiles/conclusion.tex}

{\footnotesize \bibliographystyle{acm}
\bibliography{MyFiles/mybib.bib}}
%\balance

%\theendnotes

\appendix
\input{MyFiles/appendix}

\end{document}

%% file: MyFiles/abstract.tex
The Tor anonymity system provides online privacy for millions of users, but it is slower than typical web browsing. 
%Several path selection techniques have been proposed in order to enhance the performance of Tor and attract a larger anonymity set.  %Of the current path selection proposals, the state of the art improves Tor network performance by x\%.
To improve Tor performance, we propose {\em PredicTor}, a path selection technique that uses a Random Forest classifier trained on recent measurements of Tor to predict the performance of a proposed path. If the path is predicted to be fast, the client then builds a circuit using those relays. We implemented PredicTor in the Tor source code and show through live Tor experiments and Shadow simulations that PredicTor improves Tor network performance by 11\% to 23\% compared to Vanilla Tor and by 7\% to 13\% compared to the previous state-of-the-art scheme. Our experiments show that PredicTor is the first path selection algorithm to dynamically avoid highly congested nodes during times of high congestion and avoid long-distance paths during times of low congestion. We evaluate the anonymity of PredicTor using standard entropy-based and time-to-first-compromise metrics, but these cannot capture the possibility of leakage due to the use of location in path selection. To better address this, we propose a new anonymity metric called \textit{CLASI}: Client Autonomous System Inference. CLASI is the first anonymity metric in Tor that measures an adversary's ability to infer client Autonomous Systems (ASes) by fingerprinting circuits at the network, country, and relay level. We find that CLASI shows anonymity loss for location-aware path selection algorithms, where entropy-based metrics show little to no loss of anonymity. Additionally, CLASI indicates that PredicTor has similar sender AS leakage compared to the current Tor path selection algorithm due to PredicTor building circuits that are independent of client location.

%% file: MyFiles/Intro.tex
\section{Introduction}

Privacy threats on today's Internet include targeted advertising, large-scale user profiling, and dragnet surveillance by government agencies. These threats, along with the desire to protect freedom of speech and overcome censorship on the Internet, have resulted in an increase in public interest for anonymity systems. The Tor Network~\cite{dingledine2004tor} in particular has received enormous attention and currently serves millions of users from all over the world. Tor users can connect to the Internet through an encrypted tunnel by first building a path through three Tor routers (called a \textit{circuit}) chosen from a set of approximately 7,000 volunteer routers. Part of Tor's anonymity is attributed to the size of the user base, called the \textit{anonymity set}, and attracting a large anonymity set is thus important for privacy of Tor users.  

\paragraphX{Performance.} Unfortunately, Tor is slower than typical web browsing. Several groups have proposed new circuit-building approaches that aim to improve performance by optimizing properties such as bandwidth or latency. Wacek et al.~\cite{wacek2013empirical} examined these approaches and determined 
%that performance results were poor when only latency was considered. Additionally, they determined 
that \textit{Congestion-Aware Routing (CAR)}~\cite{wang2012congestion} offered the best performance-anonymity trade-off. CAR is a decentralized approach, where clients opportunistically measure circuit congestion during circuit creation and select the best one for use. This decentralized approach is limited because clients only have a small subset of relevant congestion information. Global knowledge of congestion in Tor and performance of circuits more generally would enable better choices for all clients.
%knowledge of congestion characteristics for a small subset of routers that are measured during circuit creation. 
%Additionally, Wang et al.~\cite{wang2012congestion} concluded that long-term congestion is a property of the Tor router itself, and thus some routers are consistently more congested than others. Therefore, we believe that Tor performance can be improved if clients have global knowledge of long-term congestion characteristics of Tor routers during circuit creation.

Building on this insight, we propose \textit{PredicTor}, a path selection technique that leverages performance measurements of many circuits to
%avoid consistently congested nodes and chooses 
select less congested nodes and geographically shorter paths with greater probability. PredicTor uses a Random Forest classifier trained on recent measurements of Tor circuits to predict the performance of a proposed path. If the path is predicted to be fast, then a circuit is built using those relays. We implemented PredicTor in the Tor source code and show through simulations in Shadow that PredicTor improves Tor network performance by 23\% compared to Vanilla Tor and by 13\% compared to CAR.  This resulted in a speed up over Vanilla of over 500ms in the median case, and over 1.5s in the 90th percentile. 

Moreover, we performed live Tor experiments and show that PredicTor improves network performance partly due to avoiding highly congested nodes, and partly due to building lower latency circuits.  
%During times of high congestion, PredicTor strives more towards avoiding highly congested nodes. During times of low congestion, when there are fewer congested nodes to avoid, PredicTor strives more towards building shorter circuits. Thus, 
In fact, PredicTor is the first path selection algorithm that 
dynamically considers both congestion and latency according to the state of the live Tor Network. In the live Tor experiments, during times of high congestion, we show an improvement of 7\%-13\% in the median case for PredicTor compared to Vanilla Tor. 

%While many of these approaches improve security or performance on some level, it is crucial to understand their anonymity impact on all levels. 

%An ideal design should balance between efficiency and anonymity, and it is better to have some counter-examples that some efficient Tor designs but lead to privacy leakage. However, current anonymity metrics cannot fulfill the task, which links to CLASI that fills the gap.

%Ever since Tor was deployed in 2003, there have been proposals for new path selection algorithms to improve performance and security. In terms of performance, some proposals make optimizations based on relay bandwidth~\cite{snader2008tune}, latency~\cite{akhoondi2012lastor,sherr2009scalable}, and relay utilization~\cite{geddes2016abra}. Other proposals strive to improve security by routing around network-level adversaries to avoid traffic analysis attacks~\cite{edman2009awareness,starov2015measuring,barton2016denasa,johnson2015avoiding} or to avoid active BGP hijacking attacks~\cite{sun2015raptor,sun2017counter}. While many of these approaches improve security or performance on some level, it is crucial to understand their anonymity impact on all levels. For this reason, multiple metrics have been proposed to measure anonymity of path selection algorithms in Tor.

%For this reason, measuring anonymity of path selection algorithms in Tor has been an area widely studied by researchers.  

%\paragraphX{Anonymity.} 

\paragraphX{Measuring Anonymity.} %\paragraphX{Improving Tor}  
%\todo{disjointed transition! See my email for a coherent topic} 
Any proposal for building efficient Tor circuits must thoroughly evaluate anonymity. For example, a method that focuses on using high-bandwidth relays could concentrate traffic into fewer nodes, making it easier for a few attackers to compromise more circuits. Unfortunately, existing metrics do not address all aspects of Tor that need to be considered in evaluating new path selection proposals.

Current anonymity metrics fall into two complimentary categories: methods that aim to quantify anonymity and metrics that empirically measure all-or-nothing compromises. Most metrics that quantify anonymity are based on entropy~\cite{serjantov2002towards,diaz2003towards,rochet2016waterfiling}. %While entropy provides a useful overview of some anonymity
Syverson~\cite{syverson2009m} points out that while entropy-based metrics represent the average case well, they do not represent worst-case scenarios well.  %Meiser et al.~\cite{backes2016your} presented an algorithmic method for establishing tight upper bounds on anonymity.  In their analysis, they showed anonymity impact for two senders connecting to two receivers over an idealized Tor network. However, their analysis does not show anonymity impact for different user models or client models.  We need a metric that measures an upper bound network location leakage for Tor clients that can be used to test various path selection algorithms under various client models and user models. 
Other quantification methods~\cite{backes2016your,backes2013anoa} perform information theoretic inferences about Tor clients %by fingerprinting Tor circuits at the relay level.  %This is known as relay fingerprinting.  %In this work, we go beyond the relay level and extract features for each relay that include information about the network level, country level, and other auxiliary information such as bandwidth -- with which we use to make inferences about clients' network location.  
%All proposed fingerprinting methods seek to reveal information at the client level by making inferences 
with probability equal to \(1/|N| + \delta\) where \(N\) is the number of clients, and \(\delta\) is the degree to which the inference is successful beyond a best guess. Due to the large user base in Tor, these inference probabilities can be minuscule.  Therefore, it is difficult to justify how these inference probabilities may indicate an advantage for an adversary to fully compromise anonymity.  

The latter category, metrics that empirically measure all-or-nothing compromises, includes time-to-first compromise, a measure of how long it takes until a client uses a compromised circuit~\cite{johnson2013users}.
%inherently ignore reductions of anonymity that do not fully deanonymize a user.
%, even though these can be important.
%exclusively measure the worst case, and indicate perfect anonymity for users that have not been fully compromised. 
Though such metrics give us a good understanding of properties that lead to full deanonymization, they offer less insight into the state of anonymity for users that have not been fully deanonymized. As such, we need a metric that offers some insight into the state of anonymity before full deanonymization and one that shows an adversary's ability to infer key attributes about the user.

In this work, we present an anonymity metric called \textit{CLASI} (Client AS Inference). CLASI measures an all-knowing adversary's ability to infer clients' Autonomous Systems (ASes) by fingerprinting their circuits at the network and country level, along with other auxiliary information such as relay bandwidth. We give our adversary full knowledge of all the connections in the Tor Network, and our results thus represent an upper bound. Information revealed about the clients' AS, rather than the client directly, could potentially be more useful for adversaries for several reasons: 
\squishlist
\item The number of popular Tor client ASes is far lower than the number of Tor clients. 
%This becomes more true over time as the number of Tor clients grows faster then the number of ASes. Therefore, 
Thus, inferring a client's AS is more achievable and may be a first step in reconnaissance for an adversary;
\item High-resource adversaries such as nation-states are known to target ASes for infiltration in efforts to passively observe network traffic;
\item Making inferences at the client level may yield negligible results due to \(Pr[1/|N| + \delta]\) being small in most cases, especially when \(N\) is large.
\squishend

We evaluate this method empirically by testing a recently proposed \textit{location-aware} algorithm called DeNASA~\cite{barton2016denasa}. Comparing DeNASA to Vanilla Tor, we find anonymity loss using CLASI that is not apparent when using entropy-based metrics. We note that DeNASA is not a performance-based path selection algorithm, but rather that it seeks to improve security by routing around network-level adversaries to avoid traffic analysis attacks. Thus, CLASI can be useful for evaluating other such algorithms~\cite{edman2009awareness,starov2015measuring,barton2016denasa,johnson2015avoiding,imani2018guardsets} and for schemes that seek to avoid active BGP hijacking attacks~\cite{sun2015raptor,sun2017counter}. 

%We show the anonymity impact for DeNASA compared to Vanilla Tor using the CLASI metric and entropy-based metrics, and we find 
%. Our results indicate that CLASI shows anonymity loss for location-aware path selection algorithms, where entropy-based methods show little to no loss of anonymity. 
Finally, we evaluate the anonymity of \textit{PredicTor} using both CLASI and entropy-based metrics. We find that AS leakage for PredicTor is similar to Vanilla and slightly better than CAR due to PredicTor clients building paths independently of their own network location.

\paragraphX{Contributions} In summary, we make the following contributions:

\begin{smenum}
\item We show circuit classification accuracy for machine learning algorithms that are trained using data currently available from the Tor consensus files.

\item We present PredicTor, implement it in the Tor source code and show significant performance benefits in Shadow simulations.

\item We perform live Tor experiments and show that PredicTor is the first path selection algorithm to dynamically optimize for congestion and path length depending on path conditions. 

\item We present the CLASI anonymity metric. Our evaluation shows that CLASI indicates anonymity loss for location-aware path selection algorithms where entropy-based metrics show little to no loss of anonymity.

\item We evaluate PredicTor with CLASI and other metrics and find that PredicTor's path selection maintains high anonymity.
\end{smenum}

%The rest of the paper is organized as follows. In Section 2, we present background knowledge and related work. Section 3 motivates the PredicTor design by presenting path classification results, and we build on this to present the PredicTor design in Section 4. The CLASI design is presented and evaluated in Sections 5 and 6, respectively.  The anonymity implications for PredicTor are evaluated in Section 7. We then discuss both PredicTor and CLASI results and deployment ideas along with future work ideas in Section 8.  Finally, we conclude in Section 9.

%% file: MyFiles/background.tex
\section{Background and Related Work}
Tor is a low-latency anonymity system for TCP-based applications~\cite{dingledine2004tor}. The Tor network comprises approximately 7000 volunteer-operated relays~\cite{TorMetrics} that are deployed throughout the world. It was recently shown by Jansen et al.~\cite{jansen2016safe} that Tor has approximately 550,000 active users at any given time. Each client selects a three-hop path of relays and builds a multi-hop encrypted tunnel, called a {\em circuit}, through this path. The first, middle, and last relays on the circuit are called the \textit{guard}, \textit{middle}, and \textit{exit} relays, respectively. 
%The circuit is built such that each relay knows only its predecessor and successor. 
%Circuits are built based on the onion routing protocol, where clients negotiate session keys incrementally with each successive hop in the path until the final hop is reached.  Then, the TCP connection from the final hop to the destination remains unencrypted, unless some form of encryption is used at the application layer such as HTTPS.

A client uses a single guard node as the first hop for all of its circuits to help prevent attacks such as the \textit{predecessor attack}~\cite{wright2003defending,wright2008passive,overlier2006locating}, the selective denial of service attack~\cite{borisov2007denial}, and statistical profiling. 
%This reduces the risk that an attacker will control the first hop on a user's circuits. 
A new guard is chosen only if the presently selected guard becomes unavailable, or if a period of 60 days to 9 months is reached~\cite{dingledine2014one}.
%thereby reducing the probability of rotating to a compromised guard~\cite{johnson2013users}.

To provide fast connections for web browsing, relays are selected for circuits such that traffic is evenly distributed over the available bandwidth in the Tor Network. A set of \textit{directory servers} are responsible for securely maintaining the list of relays, along with their bandwidths and other information. Once per hour, each client receives a \textit{consensus document} from the directory servers, and this document contains weights assigned to each relay based on the relay's position in the circuit and its bandwidth. Then, load balancing is achieved by selecting each relay in proportion to its consensus weights.  

\subsection{Improving Network Performance}
%Due to congestion and latency, the Tor Network continues to be bandwidth starved.  

Tor is slower than typical web browsing, and a number of research groups have attempted to address this~\cite{snader2008tune,sherr2009scalable,akhoondi2012lastor}.
%Improving network performance in Tor is integral in attracting more users, and thereby increasing the anonymity set. To this end, there has been much work done in the space of relay selection for optimizing network performance~\cite{snader2008tune,sherr2009scalable,akhoondi2012lastor,geddes2016abra}.  
Wacek et al.~\cite{wacek2013empirical} examined these approaches and determined that \textit{Congestion-Aware Routing} (CAR)~\cite{wang2012congestion} offered the best performance-anonymity trade-off. In this paper, we thus use CAR as a benchmark for comparison.

%The state-of-the-art method, proposed by Wang et. al., is called \textit{Congestion-Aware Routing} (CAR)~\cite{wang2012congestion}, and we use it use as a benchmark in this paper to compare against our method empirically.

CAR aims to intelligently select Tor circuits with the lowest levels of congestion. Congestion measurements for circuits are performed by the clients by sampling round-trip times (RTTs) of both circuit-building and application connections. A circuit is selected for use only if it's measured {\em congestion time} (the current RTT minus the shortest RTT) is the lowest out of three randomly selected circuits. If at any point during the life of that circuit, the mean of the last five congestion times is greater than 0.5 seconds, the client will switch to another circuit.  

% mkw -- cut this, as we don't use long-term information
%An important finding by Wang et al.~\cite{wang2012congestion} was evidence that congestion is a property of the Tor router itself. Though congestion comes in bursts in the short term, each node's congestion characteristics do persist over time, and thus some nodes are consistently more congested than others while other nodes are consistently less congested. In our PredicTor approach, we aim to leverage historical circuit download times in an effort to avoid consistently congested nodes and choose consistently non-congested nodes with greater probability.
%\todo{this last bit seems to not fit the features we use. We don't (and can't) target individual nodes. If there is a congested node with good BW, CC, and AS, then we will use it.}

%Congestion Aware:
%How does it work?  How much relative improvement compared to Vanilla? What are disadvantages and advantages?

%Bandwidth starved.  Anonymity Set.

\subsection{Measuring Anonymity}

The existing literature provides substantial contributions in measuring Tor's anonymity~\cite{serjantov2002towards,diaz2003towards,snader2008tune,rochet2016waterfiling,backes2016your}. Our approach, CLASI, builds on the AnoA framework proposed by Backes et al.~\cite{backes2013anoa} for computing quantitative bounds on the anonymity in Tor. The AnoA framework is modeled as a challenge-response game between an adversary and a challenger. The adversary possesses two tables (\(D_0~and~D_1\)) in which each line is populated with a sender, a receiver, and auxiliary information. The two tables differ in exactly one row in the sender field. For this special row, the sender field for \(D_0\) contains sender \(S_0\), and the sender field for \(D_1\) contains sender \(S_1\). The adversary \(A\) sends tables (\(D_0~and~D_1\)) to a challenger \(CH\). The challenger chooses \(D_b\) according to its input \(b\) where \(b\subseteq \{0,1\}\), and successively feeds each row to an idealized Tor protocol. At any point, the adversary outputs their decision \(b\). Sender anonymity for this protocol is then measured in terms of \(\delta\) where:
\begin{displaymath}
Pr[b=0:0\leftarrow A,CH(0) ] \leq Pr[b=0:0\leftarrow A,CH(1)] + \delta.
\end{displaymath}
\noindent As anonymity of the protocol decreases, \(\delta\) increases due to the fact that adversary \(A\) guesses \(b\) correctly with greater probability. 

We use a framework similar to AnoA as the foundation for designing our CLASI metric. The most important and distinguishing characteristics of the CLASI metric are:

\squishlist
\item{We equip the adversary with a probabilistic classification model trained on realistic Tor simulated data.}
%in which each relay in the training data is resolved into AS, CC, and BW features.}
\item{Our adversary is all-knowing, and thus our metric provides an upper-bound.}
\item{Our adversary classification model is configured to infer the Autonomous System of the client.}
\squishend

%\todo{define the features.Armon: Done. MW: Always before first used}
%For the \textit{AS features}, we use the AS number directly as an integer. For the \textit{CC features}, we use the decimal representation in ASCII of the two character country code. For the \textit{BW features}, the consensus bandwidth is used and represented as an integer.  
In the CLASI classification model, we use three features for each relay in a circuit: the bandwidth of the relay from the consensus file (BW), and the network (AS) and country (CC) that the relay is located in. We decided to use AS, CC, and BW features because the proposed path selection algorithms in Tor are generally designed to optimize performance or security based on relay bandwidth~\cite{snader2008tune}, network location of relays~\cite{akhoondi2012lastor,sherr2009scalable}, or by routing around relays that located in certain ASes~\cite{edman2009awareness,starov2015measuring,barton2016denasa,johnson2015avoiding}. We measure an all-knowing adversary's ability to infer clients' ASes because knowledge of the clients' AS is a probable first step for adversary reconnaissance. The CLASI design and evaluation are described in Sections~\ref{CLASI} and~\ref{CLASI_Eval}, respectively.

\subsection{Related Work}\label{related-work}
%In this section, we briefly introduce routing protocols that have shown significant improvement in Tor network performance and anonymity metrics that have been utilized in past work for measuring anonymity of routing protocols in Tor.   

%\subsubsection*{\textbf{Routing Protocols}}
\paragraphX{Routing Protocols.}
Snader and Borisov~\cite{snader2008tune} proposed a change to Tor's path selection algorithm that allows the client to tune the degree to which relay selection is weighted in proportion to bandwidth. The tunable parameter can be increased to bias relay selection in favor of high bandwidth relays or decreased to reduce that bias and induce more uniform relay selection. A limitation of this approach is that selecting relays weighted too heavily towards bandwidth can cause high-bandwidth relays to become overloaded and low-bandwidth relays to become starved, resulting in poor performance. 

Sherr et al.~\cite{sherr2009scalable} proposed a latency-aware relay selection strategy in which relays participate in a virtual coordinate embedding system. Clients then estimate the latencies of anonymous circuits by summing the virtual distances between relays' advertised coordinates. Akhoondi et al.~\cite{akhoondi2012lastor} proposed an approach that aims to reduce latency of paths by accounting for inferred locations of relays while choosing paths. Some limitations to these approaches were pointed out by 
Wacek et al.~\cite{wacek2013empirical}, who performed an empirical study in which they compared the routing protocols mentioned above. Their results indicate that relay selection algorithms perform best when bandwidth is considered as a factor. Moreover, CAR was shown to perform close to the best in throughput and time-to-first-byte, in addition to significantly outperforming other algorithms in anonymity.   

One important disadvantage of CAR is that circuit-RTTs can be manipulated during circuit creation by malicious exit nodes. This disadvantage is compounded in another similar approach called Navigator~\cite{annessi2016navigator}, in which active RTT measurements and a-priori information from the distribution of globally measured RTT values are used to select circuits. Additionally, Geddes et. al~\cite{geddes2013low} suggested that the use of RTT measurements for latency improvements also results in an increase in the effectiveness of latency-based attacks.  

More recently, Geddes et al.~\cite{geddes2016abra} proposed ABRA (the avoiding bottlenecks relay algorithm). Their approach aims to increase network utilization by having relays estimate the extent to which they are a bottleneck on each circuit and spread this information to clients.
%spread information to clients about which 
%each relay compute a weight for each circuit that indicates 
%which for circuits they are bottlenecks on thus calculating a respective weight. %Then, similar to DWC routing~\cite{yang2001quality}\todo{what is DWC? Armon: I cited the DWC paper. MW: not sufficient. best to just cut, not critical}, 
%Then these weights are gossiped to clients, allowing clients to select circuits more intelligently. 
They showed that ABRA results had better network utilization compared to CAR. However, they did not show results for time-to-first-byte or time-to-last-byte measurements, so there is no evidence that ABRA offers any improvement in these measures of end-user performance.%\todo{did they show these results or leave them out? would be good to be clear. AB: they left these results out.} 

%\todo{need to have a clear statement about why these approaches are insufficient.  Armon: In paragraph 3 stated Wacek et al.'s findings.  I also added a limitation under SB's description. }

\paragraphX{Anonymity Metrics.} Existing anonymity metrics for Tor can be categorized into works that use information theoretic or rigorous methods to quantify anonymity of Tor users and works that aim to empirically measure all-or-nothing compromises of Tor users. Our proposed anonymity metric lies within the former category. 

In the area of quantifying anonymity, Serjantov and Danezis~\cite{serjantov2002towards} and Diaz et al.~\cite{diaz2003towards} propose using Shannon entropy~\cite{shannon2001mathematical} to measure the uncertainty of the distribution of guard/exit pairs selected by senders. %Diaz et al. propose normalizing Shannon's entropy such that maximal and minimal anonymity is represented by a degree of 1 and 0, respectively. 
Rochet et al.~\cite{rochet2016waterfiling} proposed a metric based on {\em guessing entropy} that indicates the expected number of nodes that must be compromised in order to mount a successful correlation attack. Snader and Borisov~\cite{snader2008tune} apply the Gini coefficient to measure the equality of selection probability for Tor relays. A Gini coefficient of 0 means all relays were chosen with equal frequency (maximal anonymity), and a coefficient of 1 means the same relay was always chosen (minimal anonymity). 

One limitation for entropy based metrics -- pointed out by Syverson~\cite{syverson2009m} -- is that the results can be misleading because the worst case is not always represented. Additionally, entropy does not indicate a loss in anonymity if clients select relays differently, as long as the distribution of selected relays is near uniform. To consider an extreme example, suppose client A always selects relay $X$ and client B always selects relay $Z$; the entropy would be \(1\). This is a misleading result in terms of anonymity because both clients are fully identifiable with knowledge of their selected relay.  %For our CLASI metric, the worst cases are represented because we model an all-knowing adversary, and thus, our metric provides an upper-bound.

To establish tight upper bounds on anonymity, Meiser et al.~\cite{backes2016your} presented a rigorous methodology for quantifying anonymity of Tor with respect to budget adversaries. In their analysis, they show anonymity impact for a system with two senders connecting to two receivers using several proposed path selection algorithms over an idealized Tor network. Their analysis, however, does not show anonymity impact for users who are masked within large anonymity sets or for varying user destinations. In our proposed metric, these parameters are tunable, allowing researchers to understand anonymity impact for different client models and different user models.  %Our motivation for this is that in live Tor, there are approximately 550,000~\cite{jansen2016safe} connected users at any given moment.   
%\todo{what are the limitations? why do we need new metrics?  Armon, I stated some limitations inline above.}

\begin{figure*}[t]
\vspace{-10mm}
%\vspace{-.06\textheight}
    \begin{minipage}{.3\textwidth}
    \centering\vskip -0.6cm
    \subfloat[Shadow Accuracy]{
    %\vspace{.06\textheight}
    %\vskip -1cm	
    \includegraphics[trim=0.35cm 0 1.3cm 0.3cm, clip, width=\linewidth, height=.16\textheight]{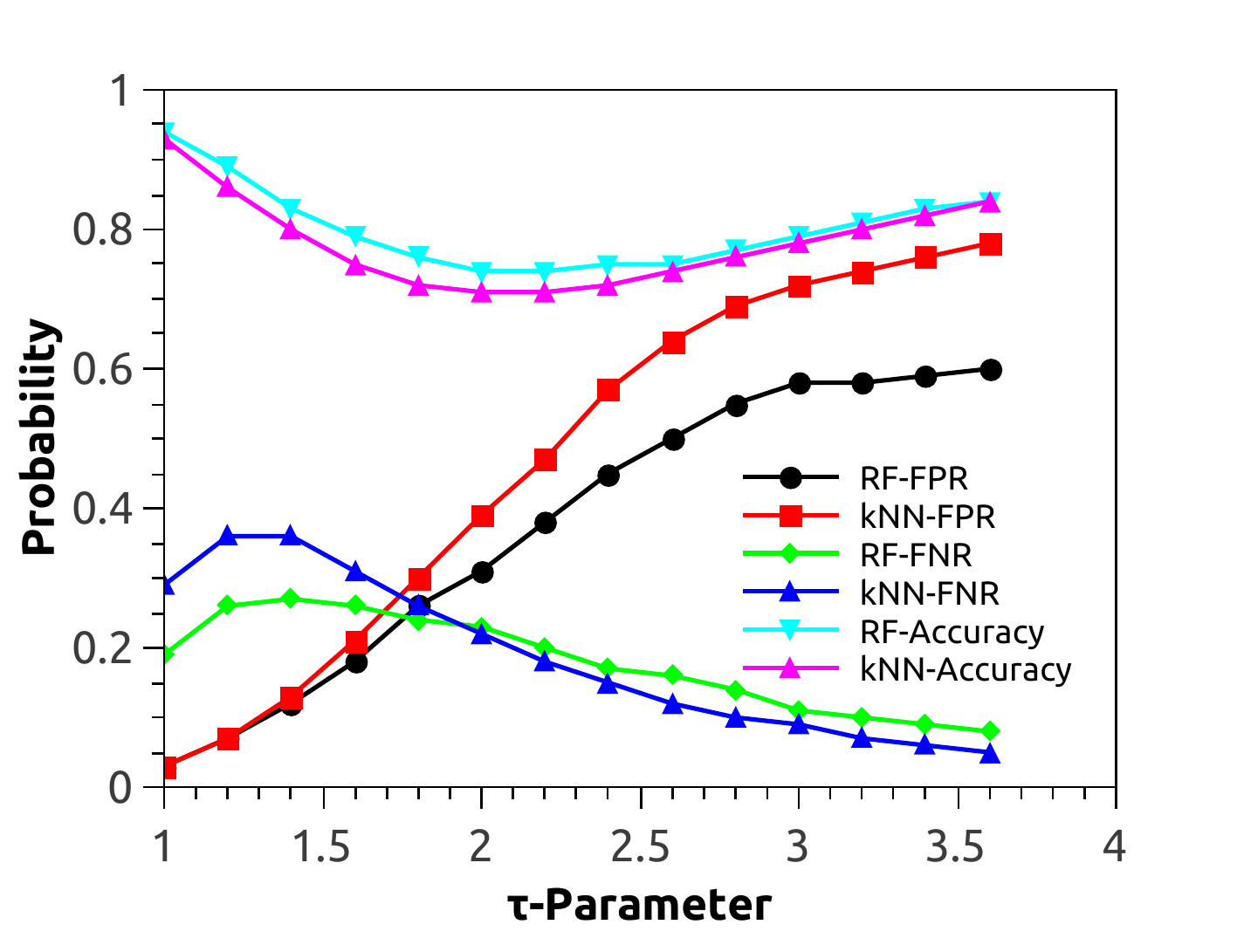}
   %\vskip -0.6cm	
    %\caption{Shadow model accuracy for All features compared to AS, CC, and BW features only.}
\label{tau-shadow}}
\vspace{-3.0mm}
%\end{figure}
\end{minipage} \hspace{0.4cm}
\begin{minipage}{.3\textwidth}
%\begin{figure}
    \centering\vskip -0.6cm
    \subfloat[Live Tor Accuracy]{
    %\vskip -1cm	
    \includegraphics[trim=0.35cm 0 1.3cm 0.2cm, clip, width=\linewidth, height=.16\textheight]{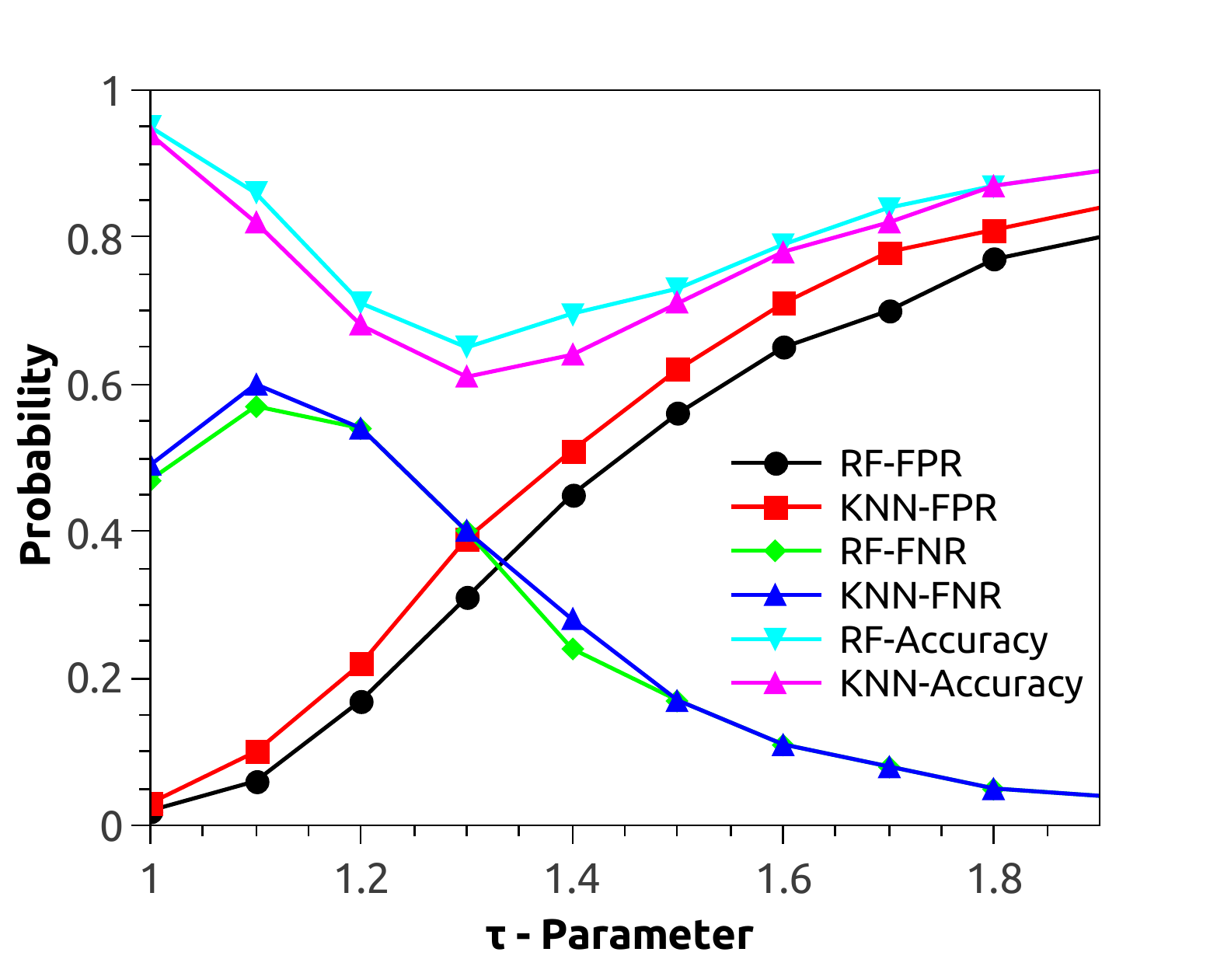}
   %\vskip -0.3cm	
    %\caption{Live Tor accuracy for All features compared to AS, CC, and BW features only}
\label{tau-liveTor}}
\vspace{-3mm}
\end{minipage}
\hspace{0.4cm}
\begin{minipage}{.3\textwidth}
%\begin{figure}
    \centering\vskip -0.6cm
    \subfloat[Live Tor Circuit Length]{
    %\vskip -1cm	
    \includegraphics[trim=0 0 0.3cm 0.2cm, clip, width=\linewidth, height=.16\textheight]{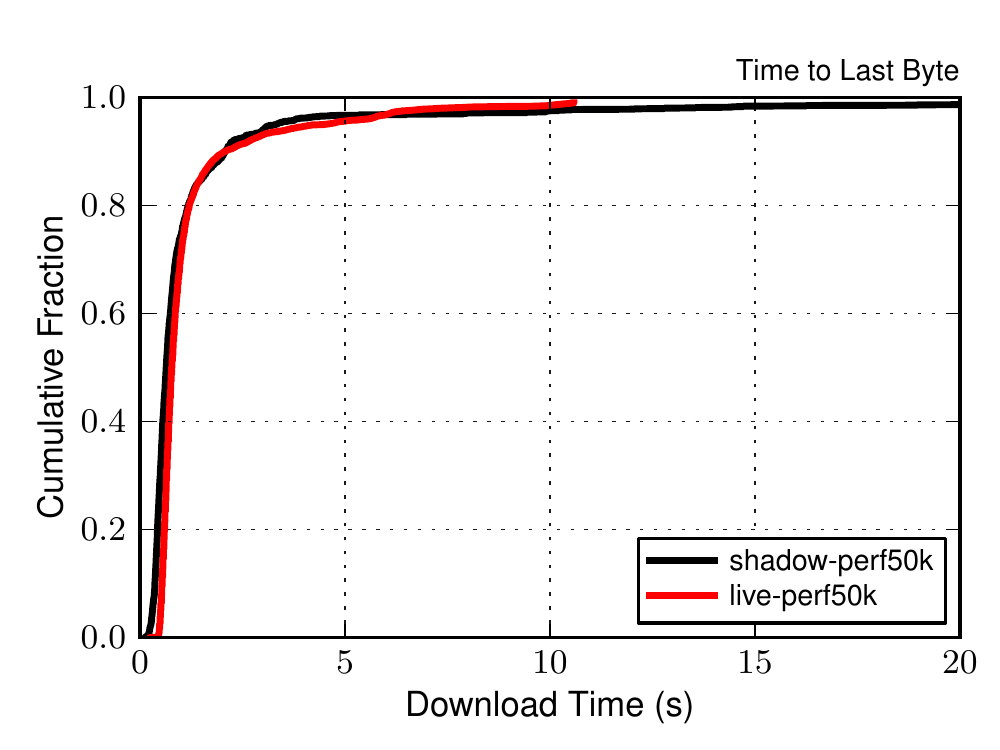}
   %\vskip -0.3cm	
    %\caption{Effect of varying \(\tau\) with respect to accuracy, false positive rate, and false negative rate.}
\label{perf}}
\vspace{-3mm}
\end{minipage}
\vspace{1mm}
\caption{Effect of varying \(\tau\) on accuracy, false positive rate, and false negative rate for $k$-NN and Random Forest in (a) Shadow and (b)  Live Tor. (c) TTLB for Shadow \textit{perf clients} compared to live Tor perf~\cite{TorMetrics} for 50KiB downloads.}\vskip -0.5cm
\end{figure*}

In the area of empirical measurement -- being complimentary to anonymity quantification metrics such as our proposal -- Johnson et al.~\cite{johnson2013users} measured time to first compromise by \textit{relay-level} and \textit{AS-level} adversaries by modeling the Tor network and taking empirical measurements. Murdoch and Watson~\cite{murdoch2008metrics} presented an analysis of proposed path selection algorithms against adversaries that deploy malicious Tor nodes. Sun et al.~\cite{sun2017counter} proposed a metric that measures the resilience of the Tor network to active attacks on BGP routing called \textit{RAPTOR} attacks.

These empirical measurement approaches are complimentary to our proposed metric because they measure all-or-nothing compromises, while our metric quantifies the ability of an all-knowing adversary to infer clients' ASes -- a property that could lead to a compromise and thus indicates a loss of anonymity for path selection algorithms under study.

%% file: MyFiles/classification.tex
\section{Path Classification}\label{path}
 
In this section, we motivate the design of PredicTor by first showing machine learning classification results for k-NN and Random Forest models trained on Tor descriptor data. Our goal is to classify Tor circuits into two classes: fast and slow. We used two distinct methods for acquiring the training data and show results for both.

\paragraphX{Shadow Data.}~\label{shadow_training} In the first method, we ran a Tor network simulation with 1000 clients using Shadow~\cite{jansen2011shadow}, a discrete-time event simulator. More details about the simulation are discussed at the end of this section. We generated a training set of 120,000 streams from one simulation run and a testing set of 25,000 streams from another simulation run. Each stream consisted of a \textit{Vanilla client} downloading a file from a server through a circuit. For each stream, we recorded the time-to-last-byte (TTLB) download time that was measured from the client during the simulation. We then set a threshold $\tau$ and labeled each data point as \textit{True} if the TTLB was less than $\tau$, i.e. the stream was fast, and \textit{False} if the TTLB was greater than $\tau$, i.e. the stream was slow.
%For all streams in the training set, the median TTLB value was \(1.8s\). Therefore, we set a threshold parameter \(\tau=1.8s\). The data point labels were then set to True and False if \(\tau < 1.8s\) and \(\tau \geq 1.8s\) respectively. 

\paragraphX{Live Tor Data.} In the second method, we gathered training data from the live Tor Network by deploying a server that hosted 20 VMs, each running Tor version 0.3.0.9. From each VM, circuits were built over the live Tor network and requests were made to download an 80 KiB file from a US destination server. For each file download, we measured the time-to-last-byte download time.  The labels were then set to \textit{True} if the TTLB was less than $\tau$, and \textit{False} if the TTLB was greater than $\tau$. Using this technique, we collected approximately 50,000 \textit{training samples} on Dec. 5, 2017 from approximately 17:00 to 18:00 GMT. Then, during the  subsequent hour (18:00 to 19:00 GMT), we collected approximately 20,000 \textit{testing samples} .

\paragraphX{Feature Set.} In Tor circuits, there is a relationship between download times and the consensus bandwidth of each relay, as well as between download times and the network location of each relay. Due to this relationship, we believe that a recognizable pattern exists such that download times can be predicted (to some degree) by inspecting bandwidth and network location of each relay in a circuit. As such, we resolve each relay into three features: 1) Autonomous System (AS), 2) Country Code (CC), and 3) Consensus Bandwidth (BW). This yields nine features for the circuit in total. For the \textit{AS features}, we use the AS number directly as an integer. For the \textit{CC features}, we use the decimal representation in ASCII of the two-character country code. For the \textit{BW features}, 
the consensus bandwidth is used and represented as an integer. We used distance-weighted $k$-NN with \(k = 9\)~\cite{mitchell1997machine}, and Random Forests~\cite{breiman2001random} to classify each circuit into class True (Fast) or False (Slow).  
 
%(\(f_1, f_4, and f_7\)), 
%(\(f_2, f_5, and f_8\)), %(\(f_3, f_6, and f_9\)), 
%More specifically, we extracted a feature set \(F\) from each circuit in our training set such that \(F = \{f_1, f_2, f_3, f_4, f_5, f_6, f_7, f_8, f_9\}\) where: \(f_1 = guard AS \), \(f_2 = guard CC \),
   %\(f_3 = guard BW \),
   %\(f_4 = middle AS \),
   %\(f_5 = middle CC \),
   %\(f_6 = middle BW \),
   %\(f_7 = exit AS \),
   %\(f_8 = exit CC \), and
   %\(f_9 = exit BW \).

\paragraphX{Classification Accuracy.}Figures~\ref{tau-shadow} and~\ref{tau-liveTor} show the accuracy, false positive rate, and false negative rate of the $k$-NN and Random Forest models in predicting circuit performance with respect to varying $\tau$. For both models, as $\tau$ increases, we observed an increase in false positive rate and a decrease in false negative rate. The false positive rate and false negative rate converge near the median download time for the training data. The median download time for Shadow and Live Tor was approximately 1.8s and 1.4s, respectively. The Shadow results in Figure~\ref{tau-shadow} show the accuracy at the median to be 76\% and 70\% for Random Forest and $k$-NN, respectively. The false positive rate and false negative rate at the median was approximately 25\% for both $k$-NN and Random Forest. The Live Tor results in Figure~\ref{tau-liveTor} show the accuracy at the median to be 70\% and 64\% for Random Forest and $k$-NN, respectively. The false positive rate and false negative rate at the median was approximately 45\% and 28\%, respectively, for both $k$-NN and Random Forest.

For both Shadow and Live Tor models, accuracy is minimal at the median download time. For greater values of $\tau$, both accuracy and false positive rate increase. Likewise, for lower values of $\tau$, both accuracy and false negative rate increase. In the context of predicting fast circuits for Tor clients, high values of $\tau$ allow clients to accept a large percentage of slow circuits due to the high false positive rate. Low values of $\tau$ cause clients to become more selective in general and lead to dramatically higher circuit build times. Based on these results, we use Random Forest for PredicTor's classification model with the $\tau$ parameter always set to the median download time with respect to the training data.  

%\begin{figure}
    %\centering
%    \vskip -0.6cm	
    %\includegraphics[width=2.8in,height=1.8in]{figs/2dplot_MLPS}
%    \vskip -0.2cm	
    %\caption{Fast and slow circuits plotted as a function of their guard consenus bandwidth vs. exit consensus bandwidth}
%\label{2d-classif}
%\vspace{-4mm}
%\end{figure}

%\paragraphX{Visualizing The Data.} 
%In Figure~\ref{2d-classif}, we plot fast and slow circuits with respect to their guard (x-axis) and exit (y-axis) consensus bandwidth weights. For visualization purposes, we zoomed in to clearly show circuits containing nine exit nodes ranging from bandwidth 116 to 129\todo{Armon added}. This data helps us see the fact that it is difficult to draw one decision boundary in 2D space that would separate the data well. This helps to explain why $k$-NN and Random Forests outperformed SVM. Another advantage of $k$-NN over SVM is the fact that SVM must be trained on a much larger training set to have high accuracy. In a real deployment, however, large training sets would result in a longer wait time for clients to receive their updated classification model. Additionally, large training sets would result in a larger consensus file. Therefore, we use a $k$-NN classification model for our PredicTor implementation, as described in Section~\ref{PredicTor}.
 
\paragraphX{Shadow Simulation Details.}~\label{shadowsim}
Our Shadow configuration consisted of 1000 clients from the top 10 countries by directly connecting users~\cite{TorMetrics}, 400 relays from a live Tor Consensus, and 70 destination servers from the Alexa list of top websites~\cite{topalexa}---forming a client to relay ratio of \(2.5:1\). All clients and relays were assigned to an enhanced network topology of 17,250 vertices and 150 Million edges based on their AS~\cite{jansennever}. In this simulation, there were two classes of clients, \textit{web clients}, and \textit{perf clients}. \textit{Web clients} randomly selected servers from which they performed HTTP GET requests to download 320 KiB files over the modeled Tor network~\cite{jansen2012methodically}, and \textit{perf clients} downloaded 50 KiB files over the Tor network. Each client measured the time from when the first request was made to when the last byte was received (\textit{TTLB}). We validated our Tor model against live Tor by comparing the results of \textit{perf clients} to historical Tor data from Tor Metrics~\cite{TorMetrics}. Figure~\ref{perf} shows the live Tor performance for fixed file size downloads of 50 KiB from historical Tor network data~\cite{TorMetrics} compared to Shadow \textit{perf clients}. The results show that live Tor performance was not significantly different than Shadow \textit{perf client} performance for our simulation, indicating that our Tor model performs statistically similar to live Tor.

%\paragraphX{Overfitting}
%\todo{Armon: added}
%We do not believe the results show characteristics of overfitting due to the excessively large circuit space compared to the minuscule training set.  For example, the Tor network model contains 400 relays and 70 servers.  More specifically, the model contains 80 guard nodes, 260 middle nodes, and 60 exit nodes.  Therefore, the number of possible streams that can be generated for this network is \(80 * 260 * 60 * 70\) = 8,7360,000 possible streams.  Our training set consisted of 125,000 streams, and our best accuracy was 79\%.  If overfitting were an issue, the accuracy would be very low due to the model no generalizing well for streams that were outside of our training set.  However, 79\% accuracy suggests that the K-NN and Random Forest models are generalizing well.  Perhaps the other models are suffering from overfitting.

%One could argue that training a model on the results of a simulation does not account for real world craziness such as cross-traffic, varying loads, etc.  To counter this argument, we point out that the accuracy was only 79\% and not near perfect accuracy.  Additionally, the accuracy for the live Tor training set was similar.

%\subsection{AS Topology}

%% file: MyFiles/predictor.tex
\section{Speeding up Tor with PredicTor}\label{PredicTor}

% \begin{algorithm}[t]
%  \KwData{\(C'\)}
%  \(c = {\emptyset}\)\;
%  \(g \leftarrow Guard List\)\;
%  c.push(g)\;
 
%  \While{True}{
  
%   \(m \leftarrow MiddleNodes\)\;
%   \(e \leftarrow ExitNodes\)\;
%   c.push(m)\;
%   c.push(e)\;
%   \(f \leftarrow E(c)\)\;
%   \eIf{\((M(C',f) == 0)\) }{
%    break\;
%    }{c.pop(e)\;
%    c.pop(m)\;
   
%   }
%  }
%  \caption{Relay selection in PredicTor}
% \end{algorithm}

We now describe PredicTor, our proposed approach for improving Tor path selection. In PredicTor, the guard selection policy is identical to Vanilla Tor, and a client will use a single guard as long as it is available for up to nine months.
%, where clients randomly select a guard from their guard list. The guard list is populated with one guard that was chosen according to its consensus bandwidth weight. Guards are only added to the guard list if the current guard's flag expires (60 days to 9 months~\cite{dingledine2014one}), or if the current guard becomes unavailable. 
To complete a path, middle and exit relays are selected according to consensus bandwidth weights as per standard Tor protocol. The resulting proposed circuit is then classified by a classification model as described in Section~\ref{path}. If the proposed circuit is predicted to be fast, the circuit is built; otherwise, new relays are selected.% (see~Algorithm~1).

Let us define function $M_{\tau}(C)$ that, for a given threshold $\tau$, returns \textit{True} when a proposed circuit $C$ is predicted to be faster than $\tau$ and \textit{False} when it is predicted to be slower than $\tau$. In the PredicTor path selection method, Tor proposes $C$ as per standard bandwidth weighted selection.  Then, if $M_{\tau}(C) == True$, the circuit is built.  Otherwise, the loop runs until the condition is met.  Note that when $\tau$ is set to be the median download time, then the loop runs two times on average, though this can vary between clients and depends on the guard selected.

\begin{figure*}[t]
\vspace{-10mm}
%\vspace{-.06\textheight}
    \begin{minipage}{.3\textwidth}
    \centering
	\vspace{-3mm}
	\subfloat[Shadow Download Times]{
    %\vspace{.06\textheight}
    %\vskip -1cm	
    \includegraphics[trim=0.38cm 0 0.38cm 0.3cm, clip, width=\linewidth, height=.16\textheight]{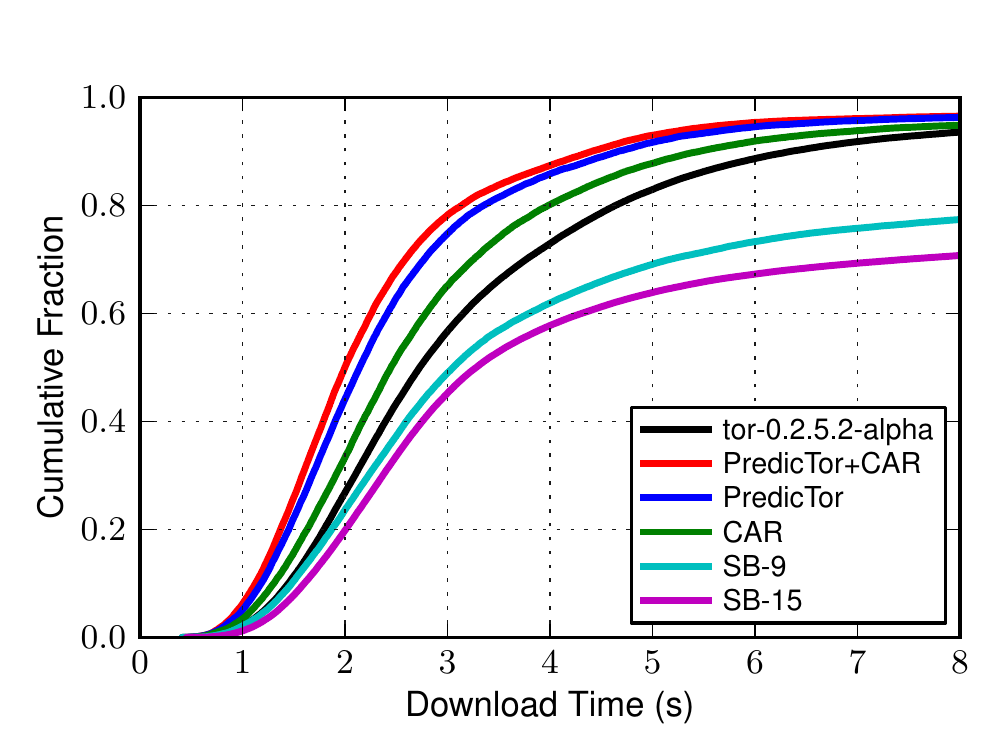}
   %\vskip -0.6cm	
    %\caption{Shadow model accuracy for All features compared to AS, CC, and BW features only.}
\label{shadow_perf}}
\vspace{-3.0mm}
%\end{figure}
\end{minipage} \hspace{.02\textwidth}
\begin{minipage}{.3\textwidth}
%\begin{figure}
    \centering
	\vspace{-3mm}
    \subfloat[Shadow Circuit Bandwidth]{
    %\vskip -1cm	
    \includegraphics[trim=0.35cm 0 0.3cm 0.2cm, clip, width=\linewidth, height=.16\textheight]{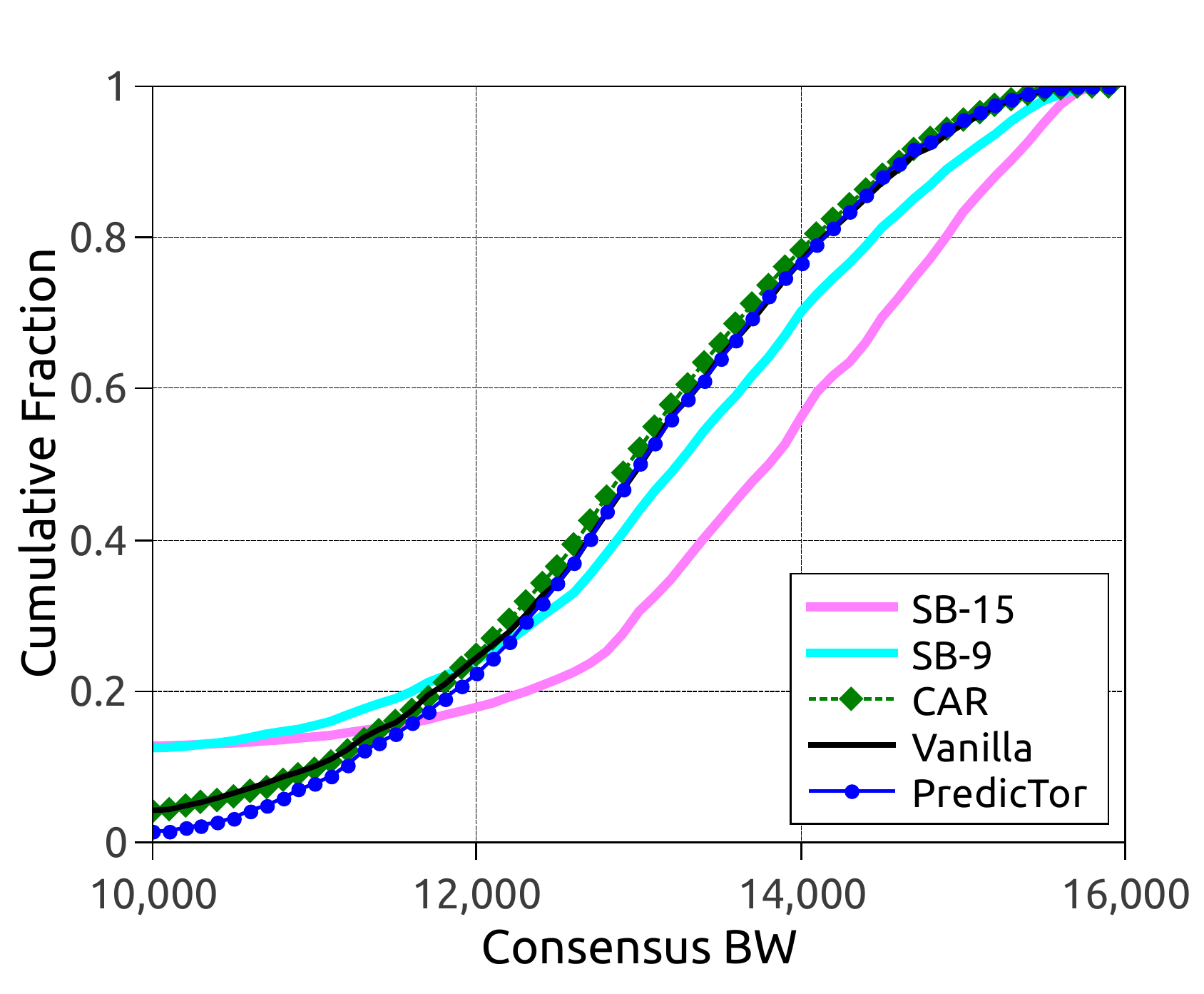}
   %\vskip -0.3cm	
    %\caption{Live Tor accuracy for All features compared to AS, CC, and BW features only}
\label{shadow_bw}}
\vspace{-3mm}
\end{minipage}\hspace{.02\textwidth}
\begin{minipage}{.3\textwidth}
%\begin{figure}
    \centering
	\vspace{-3mm}
    \subfloat[Shadow Relay Utilization]{
    %\vskip -1cm	
    \includegraphics[trim=0.25cm 0 1.46cm 0.2cm, clip, width=\linewidth, height=.16\textheight]{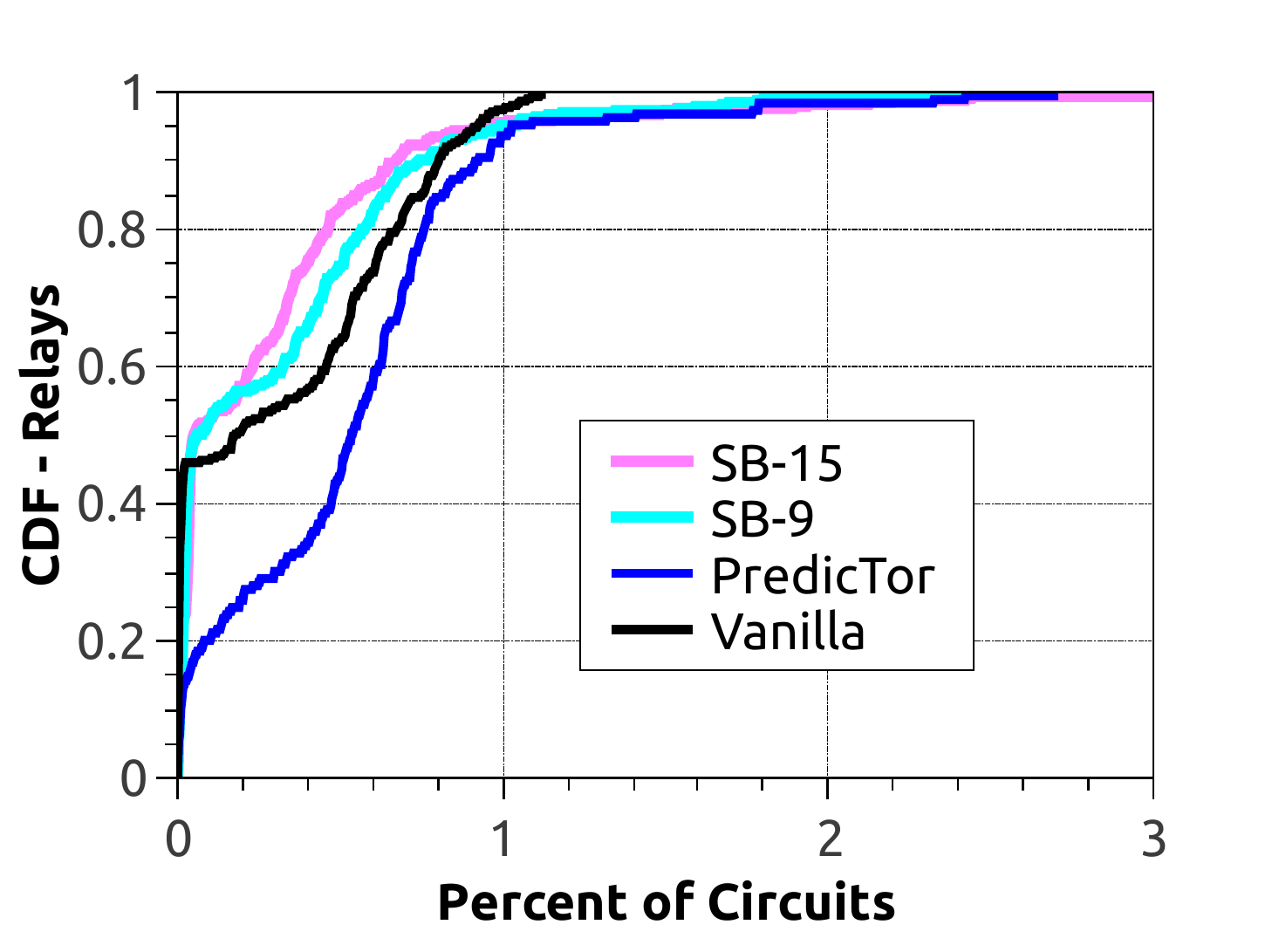}
   %\vskip -0.3cm	
    %\caption{Effect of varying \(\tau\) with respect to accuracy, false positive rate, and false negative rate.}
\label{shadow_util}}
\vspace{-3mm}
\end{minipage}
\vspace{1mm}
\caption{{\bf Shadow Experiments:} a) TTLB. b) Circuit consensus bandwidth. c) Relay utilization.  }
\vspace{-2mm}
\end{figure*}

\begin{figure*}[t]
%\vspace{-.06\textheight}
    \begin{minipage}{.3\textwidth}
    \centering
	\vspace{-3mm}
    \subfloat[Live Tor Download Times]{
    %\vspace{.06\textheight}
    %\vskip -1cm	
    \includegraphics[trim=0.3cm 0 1.4cm 0.3cm, clip, width=\linewidth, height=.16\textheight]{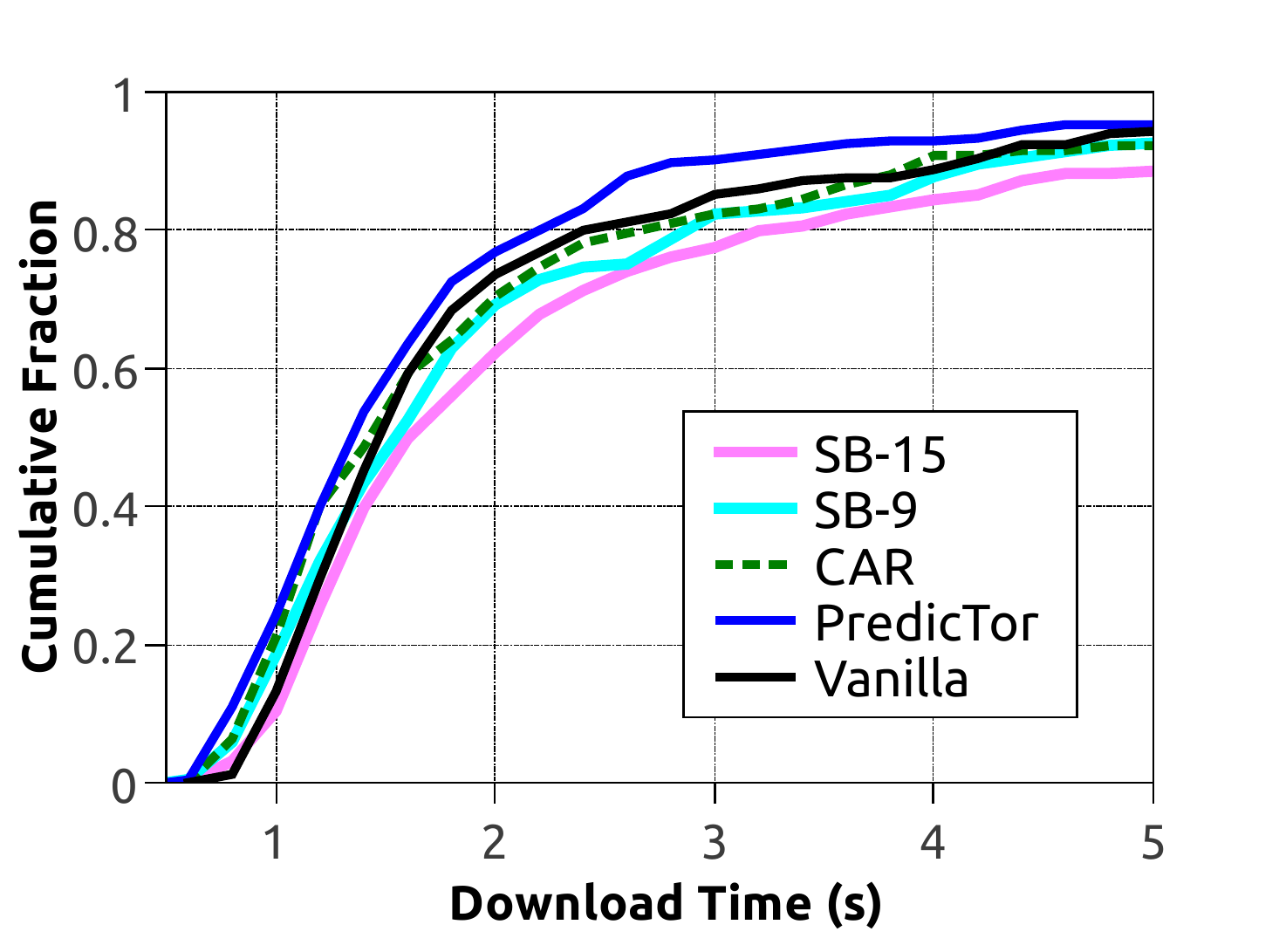}
   %\vskip -0.6cm	
    %\caption{Shadow model accuracy for All features compared to AS, CC, and BW features only.}
\label{liveTor_perf}}
\vspace{-3.0mm}
%\end{figure}
\end{minipage} \hspace{.02\textwidth}
\begin{minipage}{.3\textwidth}
%\begin{figure}
    \centering
	\vspace{-3mm}
    \subfloat[Live Tor Circuit Bandwidth]{
    %\vskip -1cm	
    \includegraphics[trim=0.3cm 0 1.9cm 0.2cm, clip, width=\linewidth, height=.16\textheight]{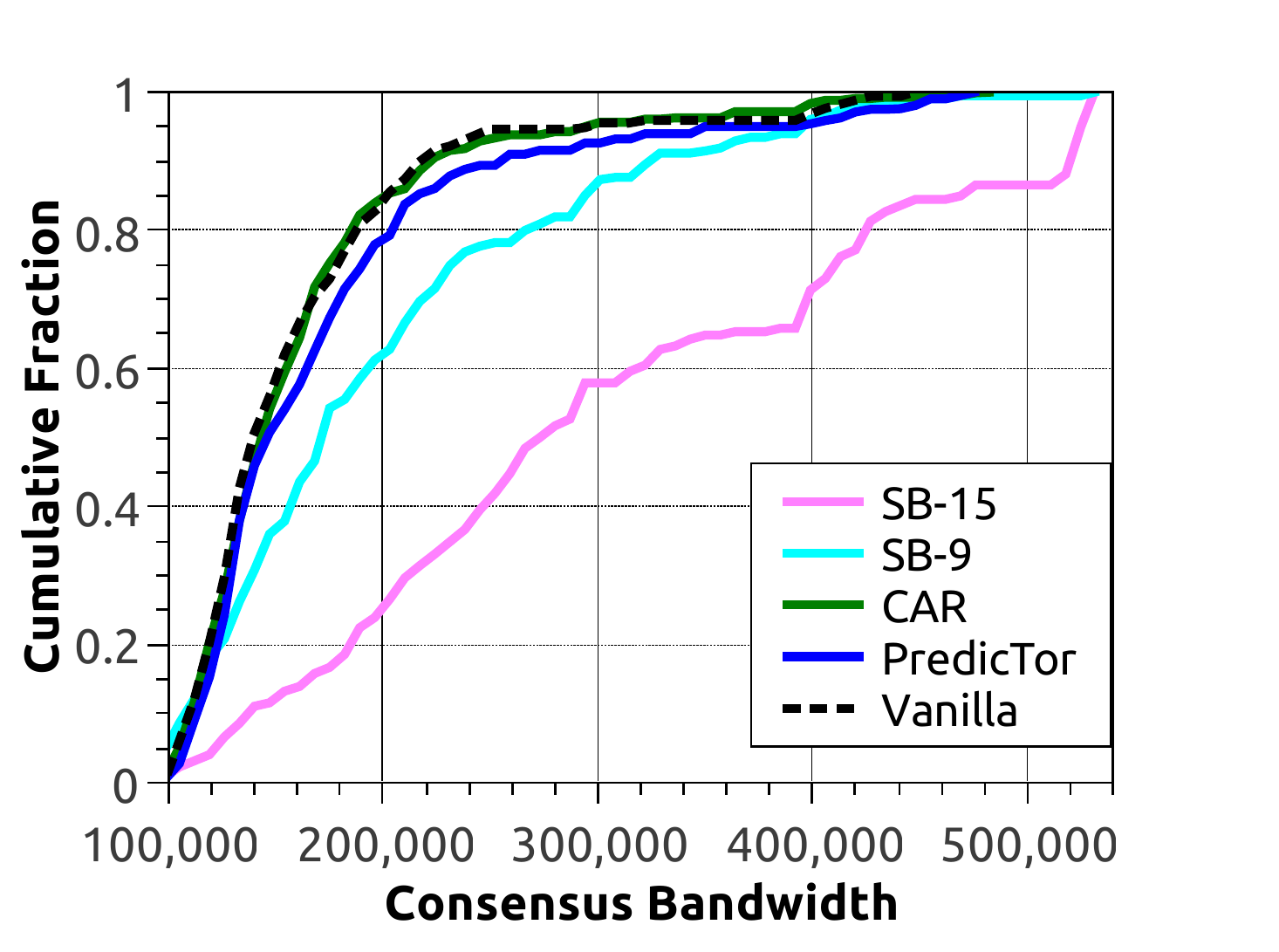}
   %\vskip -0.3cm	
    %\caption{Live Tor accuracy for All features compared to AS, CC, and BW features only}
\label{liveTor_bw}}
\vspace{-3mm}
\end{minipage}\hspace{.02\textwidth}
\begin{minipage}{.3\textwidth}
%\begin{figure}
    \centering
	\vspace{-3mm}
    \subfloat[Live Tor Circuit Length]{
    %\vskip -1cm	
    \includegraphics[trim=0.3cm 0 0.8cm 0.2cm, clip, width=\linewidth, height=.16\textheight]{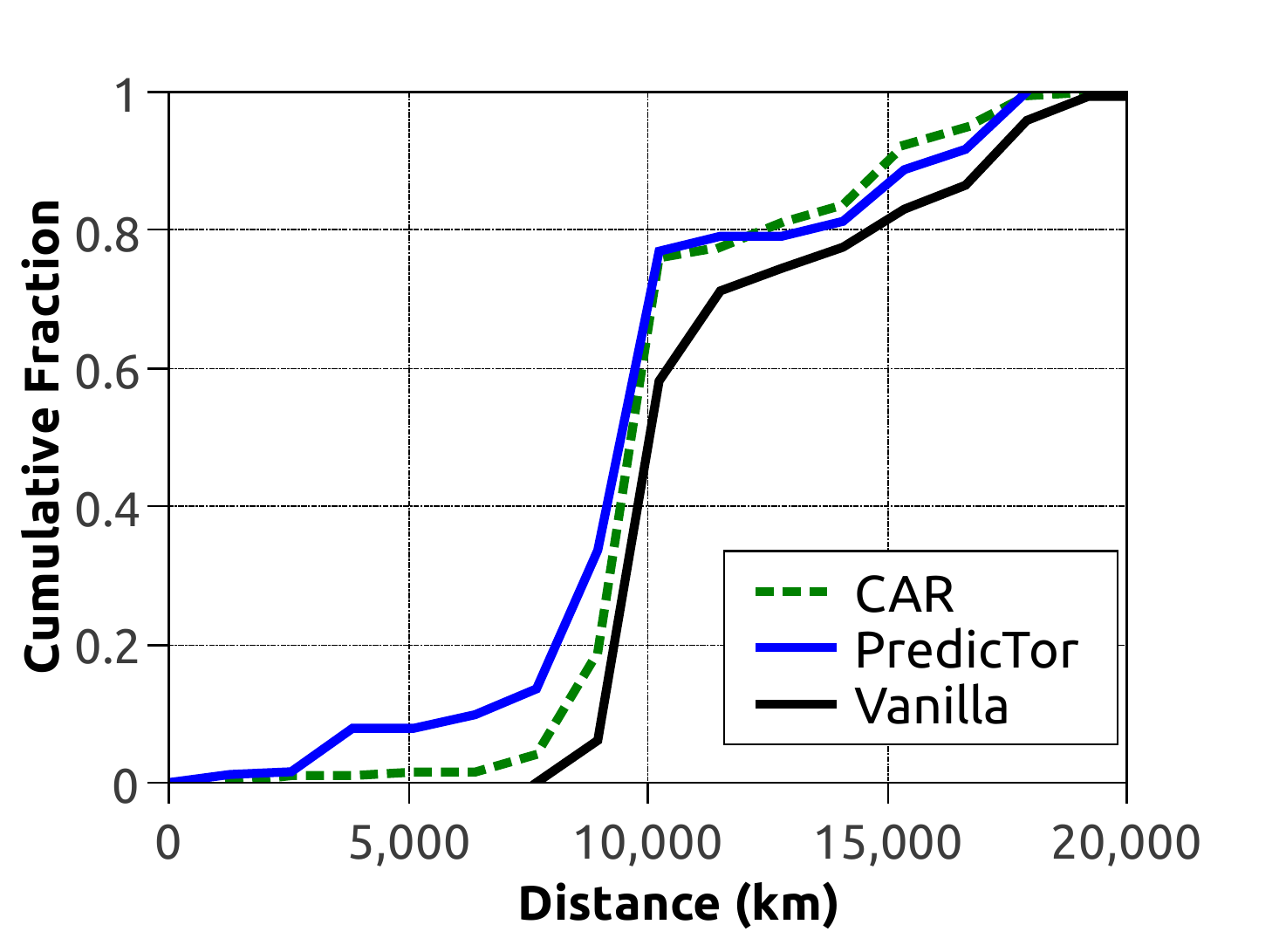}
   %\vskip -0.3cm	
    %\caption{Effect of varying \(\tau\) with respect to accuracy, false positive rate, and false negative rate.}
\label{liveTor_distances}}
\vspace{-3mm}
\end{minipage}
\vspace{1mm}
\caption{{\bf Live Tor Experiments:} a) TTLB. b) Circuit consensus bandwidth. c) Circuit length.}
\vspace{-4mm}
\end{figure*}

%Let us define two functions used in Algorithm~1. First, we have a function \(E\) that extracts features from circuits as described in Section~\ref{sec3} such that:

% \begin{displaymath}
% E:C \rightarrow F
% \end{displaymath}

% where \(C\) is the set of all possible circuits from our simulation, and \(F\) is the set of all possible feature sets.  %We denote \(c \leftarrow C\) as an event were a circuit is drawn randomly from the set of circuits.  
% \(C'\) is the set of circuits that were generated for training the classification model \(M\)~described in Section~\ref{sec3}.  \(M\) takes as input \(C'\), and maps all feature sets to \(\tau \subseteq \{0,1\}\).  More formally: 

% \begin{displaymath}
% M(C'):F \rightarrow \tau~|~\tau \subseteq \{0,1\} 
% \end{displaymath}

\paragraphX{Experimental Setup.}~\label{3ps-results}
We implemented PredicTor in the Tor source code and tested its performance compared to Vanilla, Congestion-aware routing (CAR), and Snader and Borisov (SB) path selection using both \textit{Shadow} and \textit{live Tor}.  Prior work shows that $SB$ has competitive performance under medium congestion with the parameter $s$ set to 9~\cite{wacek2013empirical,snader2008tune}. We tested $SB$ with two settings: {\em SB-9}, with $s=9$ for \textit{partial} bias to high bandwidth and {\em SB-15}, with $s=15$ for \textit{heavy} bias to high bandwidth.   

In the Shadow simulation, we used the same configuration as described in section~\ref{shadowsim}.  For all path selection techniques, the respective clients requested a 320 KiB file download from a server selected uniformly at random from a set of 70 destination servers.

In the live Tor experiments, for all path selection techniques, the respective clients requested the home page of websites selected uniformly at random from a set of 1000 sites from the Alexa list of top sites~\cite{topalexa}.

For the PredicTor experiment in both Shadow and live Tor, $\tau$ was set to the median download time with respect to the training set, and Random Forest was used for the classification model.  
Note that in a Shadow simulation, we can observe how performance is affected when \textit{all clients} use a given path selection technique.  This is not possible in live Tor because we can only deploy an insignificant fraction of clients compared to the full user base.  However, two of Shadow's limitations are: 1) the network size is significantly smaller than the real-world Tor Network, and 2) the simulation does not fully model real-world network dynamics. In live Tor, we can observe how path selection techniques respond under dynamic real-world network conditions. Therefore, for measuring performance of path selection techniques, it is useful to test in both Shadow and live Tor.   

\paragraphX{Performance Results.}   
Figures~\ref{shadow_perf} and~\ref{liveTor_perf} show Shadow and live Tor download times for Vanilla, PredicTor, CAR, SB-9, and SB-15. In both Shadow and live Tor, PredicTor was the fastest. In the Shadow simulation, PredicTor had a 23\% and 13\% median improvement compared to Vanilla and CAR, respectively, and a 28\% improvement in the 90th percentile compared to Vanilla. This resulted in a speed up over Vanilla of over 500ms in the median case, and over 1.5s in the 90th percentile. In the live Tor experiments, PredicTor had 11\% and 6\% median improvements compared to Vanilla and CAR, respectively, and a 28\% improvement in the 90th percentile compared to Vanilla. This resulted in a speed up of over 1.0 second in the 90th percentile compared to Vanilla.

\paragraphX{Circuit Bandwidth.}
SB-9 and SB-15 performed the slowest in both Shadow and live Tor. 
%In Shadow, SB-9 and SB-15 performed 8\% and 18\% slower, respectively, in the median case compared to Vanilla and 4\% and 8\% slower, respectively, compared to Vanilla in live Tor.  
Figures~\ref{shadow_bw} and~\ref{liveTor_bw} show the Shadow and live Tor circuit consensus bandwidths for Vanilla, PredicTor, SB-9, and SB-15. As expected, SB-9 and SB-15 build circuits with significantly higher bandwidth compared to other techniques, particularly in live Tor, where SB-9 and SB-15 circuits used 22\% and 97\% more bandwidth in the median than Vanilla.
%In Shadow, SB-9 and SB-15 circuits with 2\% and 6\% more consensus bandwidth in the median case compared to Vanilla, and 22\% and 97\% compared to Vanilla in live Tor respectively.  
The Shadow results suggest that selecting relays weighted heavily towards bandwidth causes high bandwidth relays to become overloaded, resulting in poor performance. Moreover, the live Tor performance and consensus bandwidth results suggest that high-bandwidth relays experience some persistent congestion due to a large user base using Vanilla's bandwidth-weighted selection policy.  The persistent congestion causes poor performance even if one client wights their selection heavily toward these high bandwidth relays.

Performance gains in PredicTor, on the other hand, cannot be attributed exclusively to selecting high-bandwidth relays. In Shadow, PredicTor did not build higher bandwidth circuits compared to Vanilla. In live Tor, Predictor uses approximately the same median consensus bandwidth as Vanilla, though it uses 25\% more bandwidth at the 90th percentile.

%\subsection{Circuit Bandwidth}
%Intuitively, picking higher-bandwidth relays should improve performance. Selecting relays weighted too heavily towards bandwidth, however, can cause high-bandwidth relays to become overloaded and low-bandwidth relays to become starved, resulting in poor performance. In Figure{}, we show a CDF of circuit consensus bandwidths for all circuits created by PredicTor compared to Vanilla and CAR during our Shadow experiment. There was a slight increase in median circuit bandwidth for CAR and PredicTor by \(11\%\) and \(8\%\), respectively, compared to Vanilla Tor. The overall distributions of bandwidth are similar among the three schemes. This indicates that the performance gains in PredicTor cannot be attributed only to differences in the bandwidths of selected nodes.

\paragraphX{Node Congestion.}
Wang et al.~\cite{wang2012congestion} concluded that congestion is a property of the Tor router itself. Though congestion comes in bursts in the short term, each node's congestion characteristics do persist over time, and thus some nodes are consistently more congested than others.  %Likewise, some nodes were found to be consistently non-congested.  
%-- centralized scheme
%PredicTor uses a centralized scheme that first measures download times over all nodes to identify consistently congested and non-congested nodes. It then distributes the data to all clients for use during future circuit creation. The advantage for PredicTor is that clients have more global knowledge of persistent network congestion for all nodes during circuit creation. This helps PredicTor clients avoid consistently congested nodes and select consistently non-congested nodes with greater probability.  

Figure~\ref{shadow_util} shows the empirical distribution (ECDF) of relays with respect to the percent of circuits that they were used on for the Shadow simulation. Vanilla completely avoided selecting approximately 45\% of relays in the network, and SB-9 and SB-15 completely avoided selecting approximately 50\% of relays in the network. Under-utilizing the network in this way caused more persistent congestion on the 65\% and 50\% of relays that Vanilla and SB did utilize, respectively. On the other hand, PredicTor utilized approximately 85\% of the network. These results suggest that, when all clients use PredicTor, more relays are utilized, resulting in greater load distribution and lower persistent congestion for high-bandwidth relays.

% respectively. 
%More details about the deployment of PredicTor are discussed in Section~\ref{disc}.

%PredicTor uses a centralized scheme that first measures download times over all nodes in a simulation to identify consistently congested and non-congested nodes, then deploys simulation data to all clients for use during future circuit creation. The advantage for PredicTor is that clients have more global knowledge of persistent network congestion for all nodes during circuit creation. This helps PredicTor clients avoid consistently congested nodes and select consistently non-congested nodes with greater probability respectively. More details about the deployment of PredicTor are discussed in Section~\ref{disc}.

%-- being a decentralized scheme --
%In CAR, on the other hand, clients only have knowledge of congestion characteristics for a small subset of nodes that are opportunistically measured during circuit creation. Therefore, in our simulation of CAR, clients did not avoid consistently congested nodes or select consistently non-congested nodes with greater probability respectively compared to PredicTor. 
%\todo{need more details/numbers here. MW: after thinking about it more, it should not be the case that PredicTor is good at avoiding consistently congested nodes. Or, if it is, then it is over-fitting relative to the features we have.}

%\todo{this last bit seems like speculation, and thus discussion section.Armon: I reference more data and discuss below.}
\paragraphX{PredicTor+CAR.}
One advantage for PredicTor compared to CAR is that clients have more global knowledge of persistent network congestion for all nodes during circuit creation. This helps PredicTor clients avoid consistently congested nodes and select consistently non-congested nodes with greater probability. In CAR, on the other hand, clients only have knowledge of congestion characteristics for a small subset of nodes that are opportunistically measured during circuit creation.

We combined PredicTor with CAR because we suspected that CAR should have better performance if nodes are selected using the PredicTor scheme first, then opportunistically measured. Figure~\ref{shadow_perf} shows a 28\% improvement in the median case for PredicTor+CAR compared to Vanilla. These results indicate that a hybridized scheme combining centralized and decentralized congestion measurements for relay selection results in better performance compared to either scheme alone.

\paragraphX{Circuit Length.}  
Although geographic distance is not a good measure for Internet latency, it can provide a point of reference for a system like Tor, where a circuit might traverse multiple intercontinental hops. Figure~\ref{liveTor_distances} shows live Tor circuit lengths for Vanilla, CAR, and PredicTor. To measure circuit lengths, we first resolved each relay into coordinates. Then, we calculated the distance between relays using Vincenty's Formula~\cite{vincenty1975direct}.  The circuit length was taken as the sum of the distances between the guard and middle and between the middle and exit. In the median case, PredicTor built circuits that were approximately 680 km shorter compared to Vanilla.  In the 90th percentile, PredicTor and CAR built circuits that were approximately 592 km and 2,043 km shorter compared to Vanilla, respectively. These results suggest that the performance gains for PredicTor and CAR are partially due to building shorter circuits, and thus, circuits of lower latency.

\subsection{Discussion}
We conclude that performance gains for PredicTor are achieved by considering three key factors: 1) congestion, 2) bandwidth, and 3) latency. From the Shadow simulation, we observed that PredicTor utilizes the network in a way that leads to more efficient load distribution and lower congestion for high-bandwidth nodes. Additionally, the live Tor results suggest that PredicTor avoids highly congested nodes while building circuits of slightly higher bandwidth and lower latency compared to Vanilla.

\paragraphX{Quantifying improvement.} Our experiments provide strong evidence that PredicTor should result in an overall improvement for all clients in Tor, with 23\% improvement in Shadow with all clients using PredicTor and 11\% improvement in the live Tor experiments with one client using PredicTor. We do not claim, however, that our experiments can show the exact quantitative gains that PredicTor would provide when deployed in Tor. One way to more fully quantify the improvement for a live deployment of PredicTor would be to test in a wide-area testbed where all clients use PredicTor. Since no such testbed was available for this study, we leave this for future work.   

\paragraphX{Malicious Relays.} We also highlight some important mitigation steps against relays that attempt to manipulate their bandwidth contribution during live PredicTor measurements to win more traffic. First, PredicTor selects guards exactly the same way as Vanilla, by consensus weight. Thus, malicious relays cannot win more guard traffic because they do not have the ability to change their consensus weight by gaming PredicTor measurements. A malicious exit relay may attempt to win more traffic by prioritizing measurement circuits and throttling all other connections, thereby appearing fast during measurement.  This can be mitigated by selecting probe destinations from the distribution of most popular destination websites as observed from (honest) exit nodes, reported safely using a system like PrivCount~\cite{jansen2016safely}. Since popularity of websites is heavily concentrated in relatively few sites~\cite{crovella1998heavy}, a moderate-sized list of probe destinations should suffice to make the attacker unable to distinguish quickly between a measurement circuit and the majority of non-measurement user activity. 
%Then, if a malicious exit prioritizes a measurement circuit, it would also need to prioritize a significant portion of their other connections.

In contrast, a major disadvantage for methods that use RTT measurements such as CAR and Navigator is that malicious exit nodes can easily manipulate RTT measurements. Geddes et. al~\cite{geddes2013low} show how the use of RTT measurements for latency improvements results in an increase in the effectiveness of latency-based attacks.  

%latency is represented in CAR's opportunistic measurements allowing CAR to build lower latent circuits.  Furthermore PredicTOr 
%that both PredicTor and CAR build lower latent circuits compared to Vanilla.

%% file: MyFiles/liveTor.tex
\section{Client Location and Guard Diversity}\label{live}
We performed additional live Tor experiments from three additional client locations: 1) United States (US), 2) Germany (DE), and 3) Japan (JP). For each client location, the experiment was performed during prime Internet surfing hours for both the US and Europe (approximately 14:00 GMT) and during a time that is evening in the US and middle of the night in Europe (approximately 00:00 GMT). We call the experiments run at 14:00 GMT as the \textit{high-congestion} condition and the experiments run at 00:00 GMT as the \textit{low-congestion} condition.

Due to the single-guard selection strategy in Tor, clients may be connected to a slow or fast guard for long periods of time. We desire to understand PredicTor performance when connected to guard nodes of various consensus weights. Thus, for each client location, the experiment was performed using a \textit{slow guard}  (consensus weight 1770) and a \textit{fast guard} (consensus weight 35600).    

In Appendix~\ref{a1}, Table~\ref{tab:1}, we show the median and 90th percentile improvement  for PredicTor compared to Vanilla. We observed that the best performance improvement from PredicTor was realized during times of high congestion while connected to a fast guard. From the US location, there was a 9.7\% and 17.7\% improvement in the median and 90th percentile, respectively. From the DE location, there was a 12.8\% and 25.3\% improvement in the median and 90th percentile, respectively. From the JP location, there was a 6.3\% and 10.8\% improvement in the median and 90th percentile, respectively.  

During times of low congestion while connected to a fast guard, PredicTor performance did not improve compared to Vanilla as much as in the high-congestion experiment. The median improvements from the US and DE were 4.2\% and 7.3\%, respectively. 
% mkw -- keeping this out for now
%For JP, the improvement in the median was 7.4\%; note that the low-congestion condition corresponds to 09:00 in Japan, so local congestion may be higher than for the high-congestion condition (23:00 Japan time).
%
%This was evident from the US and DE location, where the median improvement was 4.2\% and 7.3\% respectively.  It was also indicated 
Wang et al.~\cite{wang2012congestion} also state that CAR should get better performance during high congestion times compared to low congestion.  

We observed a slight improvement for PredicTor
%during low congestion 
while connected to a slow guard during both high and low congestion for the median time (3.3\% to 7.3\% faster). For slow guards with high congestion, PredicTor showed larger improvements for the 90th percentile, between 14.0\% and 19.1\% improvement. Slow guards typically act as a bottleneck in most circuits, but when congestion is high, PredicTor can find and select faster circuits.

%. This was evident from the US, DE, and JP location where the 90th percentile improvement was 6.3\%, 2.6\%, and 4.5\% respectively.  These results indicate that the slow guard was acting as a bottleneck in most connections.  

\paragraphX{Circuit Distance.} In Appendix~\ref{a1}, Table~\ref{tab:1}, we show the median and 90th percentile circuit distance improvements for PredicTor compared to Vanilla. We observed the best improvement in circuit distance for PredicTor during times of low congestion while connected to a fast guard. From the US location, there was a 52\% improvement in the median, with circuits that were approximately 1,600  km shorter. Similar results were observed for the DE and JP locations. %as shown in Table~\ref{tab:1}.  
During times of high congestion while connected to a fast guard, PredicTor showed more modest improvements in circuit distance of between 11\% and 26\% in the medians.
%circuit distances for PredicTor were observed to improve significantly. However, the improvement was not as great compared to the low congestion experiment.  This was evident from the US, DE, and JP location, where the median improvement was 23\%, 26\%, and 11\% respectively.  
We observed little to no improvement in circuit distance for the slow guard experiments. We believe this is due to the slow guard acting as a bottleneck in most connections.

We conclude that PredicTor intelligently picks relays in a way that has never been done by any other algorithm. During times of high congestion, PredicTor correctly avoids highly congested nodes. During times of low congestion, when there are fewer congested nodes to avoid, PredicTor correctly builds lower-latency circuits of shorter geographic distance. 

%The rest of the paper is mostly devoted to security analysis. First, we present a new anonymity metric. Then, we use that metric along with other standard metrics to analyze the security of PredicTor.  Finally, in Section~\ref{disc}, we discuss deployment ideas for PredicTor and future work ideas.

%% file: MyFiles/anonymity.tex
\section{CLASI: Client AS Inference}\label{CLASI}
Since PredicTor and other Tor path selection algorithms such as TAPS~\cite{johnson2015avoiding}, DeNASA~\cite{barton2016denasa}, and LASTor~\cite{akhoondi2012lastor} use network location information to select paths, it is important to understand the extent to which these choices lead to predictability and loss of anonymity. In particular, we seek to understand whether an attacker can infer something about the location of the client from the choices of paths that she makes. 

To this end, we now describe CLASI, a metric for measuring the ability of the attacker to infer the client AS. CLASI is a challenge-response game between an adversary and a challenger. The adversary possesses a path simulator \(PS\) that is an idealized Tor network with a given path selection algorithm to generate paths over the sender space \(S\), the relay space \(R\), and the destination space \(D\). We denote one path $P$ being generated from \(PS\) as \(P \leftarrow PS\), and a set of paths $\mathbf{P}$ being generated from \(PS\) as \(\mathbf{P} \leftarrow PS\). Each path \(P\) is a set of nodes where \(P = \{p_1,p_2,p_3,p_4,p_5\}\), such that: \(p_1~=~client IP, p_2~=~guard IP, p_3~=~middle IP,\) \(p_4~=~exit IP, p_5~=~destination IP\). 

Adversary \(A\) sends path simulator \(PS\) to the challenger \(CH\). \(CH\) generates a path \(P'\) from \(PS\) and removes the sender \(p_1'\) such that \(P' = \{p_2',p_3',p_4',p_5'\}\).  \(CH\) then sends \(P'\) to \(A\). \(A\) attempts to predict the network location \(L\) of sender \(p_1'\). More precisely, $L$ is the sender's AS, and we let $S_L$ be the set of all possible sender ASes. Then let $L'$ be $A$'s prediction for the location of the sender. Sender location information leakage for the idealized Tor network is then represented by \(\epsilon_s\), where:

\begin{displaymath}
%Pr[L' = S_L'' : S_L' \leftarrow CH, S_L'' \leftarrow A ] = \frac{1}{|S_L|} + \epsilon_s
Pr[L = L'] = \frac{1}{|S_L|} + \epsilon_s.
\end{displaymath}

There is no leakage ($\epsilon_s = 0$) if the attacker can do no better than guessing $L'$ uniformly at random from among all possible sender ASes in $S_L$. Otherwise, the attacker has some advantage in guessing the sender's AS ($\epsilon_s > 0$).

\begin{figure}
    \centering
    %\vskip -0.6cm	
    \includegraphics[width=3.1in, height=1.8in]{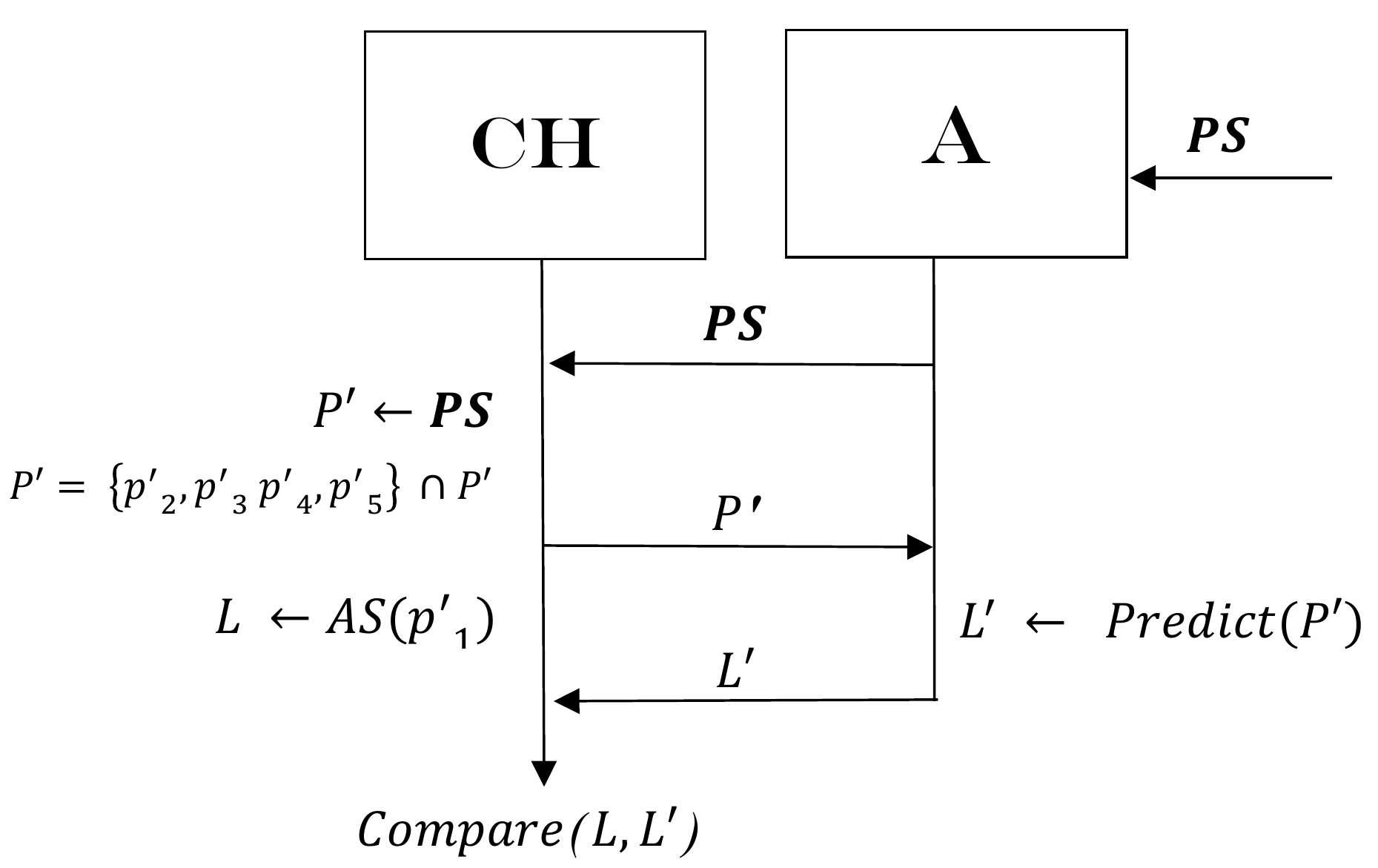}
   %\vskip -0.2cm	
    \caption{CLASI challenge-response game between adversary and challenger.}
\label{game}
\vspace{-3mm}
\end{figure}

%\todo{update the figure with new variable names, remove Extract function. Also, larger fonts.  Armon: Done}
%\todo{great. Now, I would remove the box, space CH and A out a bit, maximize use of the horizontal space}
The CLASI challenge-response game sequence is shown in Figure~\ref{game}. Adversary \(A\) uses the function $Predict(P)$ that extracts the features from the path $P$ and uses them to classify the AS of the sender. Each Tor relay has the features AS, BW, and CC, represented as described in Section~\ref{path}, while the destination has the features AS and CC. \(Predict\) uses a probabilistic classification model that is trained on the feature set of paths generated from \(PS\) and labels that represent the sender's AS.
%/ label tuple output from \(A_{Extract}\).  \(A_{Predict}\) takes as input a feature set \(F'\) and returns an AS number that represents a predicted sender's location corresponding to \(F'\) as it relates to the training data.  More specifically:

% \begin{displaymath}
% Predict(\mathbf{F},\mathbf{L}) : \mathbf{F'} \rightarrow \mathbf{S_L}
% \end{displaymath}

%two functions: 1) \(A_{Extract}\), and 2) \(A_{Predict}\). \(A_{Extract}\) resolves each path into a feature set and label; each Tor relay is resolved into its AS, BW, and CC, and the destination is resolved into its AS and CC.  The sender's AS is assigned to the label.  More specifically:

% \begin{displaymath}
% A_{Extract} : \mathbf{P} \rightarrow \mathbf{F},\mathbf{L}
% \end{displaymath}

% where
% \begin{displaymath}
% F_i=\{p^{AS}_{i2},p^{CC}_{i2},p^{BW}_{i2},p^{AS}_{i3},p^{CC}_{i3},p^{BW}_{i3},p^{AS}_{i4},p^{CC}_{i4},p^{BW}_{i4},p^{AS}_{i5},p^{CC}_{i5}\}
% \end{displaymath}

% and 
% \begin{displaymath}
% L_i = \{p^{AS}_{i1}\}
% \end{displaymath}

\section{CLASI Evaluation}\label{CLASI_Eval}

We now evaluate CLASI's ability to measure sender location information by running a simulation of the Tor network using TorPS~\cite{torps} and testing a location-aware protocol called DeNASA~\cite{barton2016denasa}.
%\todo{hopefully this will be expanded beyond DeNASA. Armon: hopefully in future work, or the next revision of this paper.}

\subsection{Tor Model}\label{torps}
TorPS~\cite{torps} is a Tor path selection simulator that uses historical data to recreate network conditions experienced by Tor users in the real world~\cite{johnson2013users}. Circuits are created according to past network state, and streams are attached to those circuits according to simulated user behavior.

For the set of destinations used in our simulation, we tested the 200 top Alexa~\cite{topalexa} websites.
%, which ranged in genre from banking, e-commerce, news, social media, search engines, blogs, and classifieds, etc.
We modeled clients connecting from the top 10 countries by directly connecting users according to Tor Metrics~\cite{TorMetrics}. The simulated clients connected from distinct ASes chosen partly from the list proposed by Edmond and Syverson~\cite{edman2009awareness}, partly from the list proposed by Juen~\cite{juen2012protecting}, and partly from CAIDA Top Ranking ASes~\cite{caidaASrank}. 

We tested four distinct users models: 1) 5-destination, 2) 10-destination, 3) 15-destination, and 4) 20-destination.  According to the number of destinations specified for each user model, clients selected their destinations uniformly at random from the set of 200 sites at the start of the simulation. During the simulation, clients connected to destinations selected uniformly at random from this pre-selected set.

\subsection{DeNASA Protocol}
Location-aware protocols are designed to increase security against AS-level threats~\cite{johnson2013users}, or the threat of BGP hijacking attacks~\cite{sun2015raptor}. For our evaluation, we chose to test a location-aware protocol called DeNASA (destination-naive AS-awareness) because DeNASA's tunable parameters allow users to increase or decrease location awareness in exchange for more or less security against AS-level adversaries respectively. We would expect, however, that 
%However, we show that 
increasing location awareness would cause an increase in sender information leakage.

DeNASA increases security against AS-level adversaries by creating circuits that have higher probability of avoiding some Tier 1 ASes from the client to guard, and simultaneously from exit to destination. Barton et al. identify eight Tier 1 ASes, called~\textit{suspect ASes} that are the most likely to appear on both sides of Tor circuits. 

\begin{figure}
    \centering
    \vskip -0.6cm	
    \includegraphics[width=2.8in, height=1.6in]{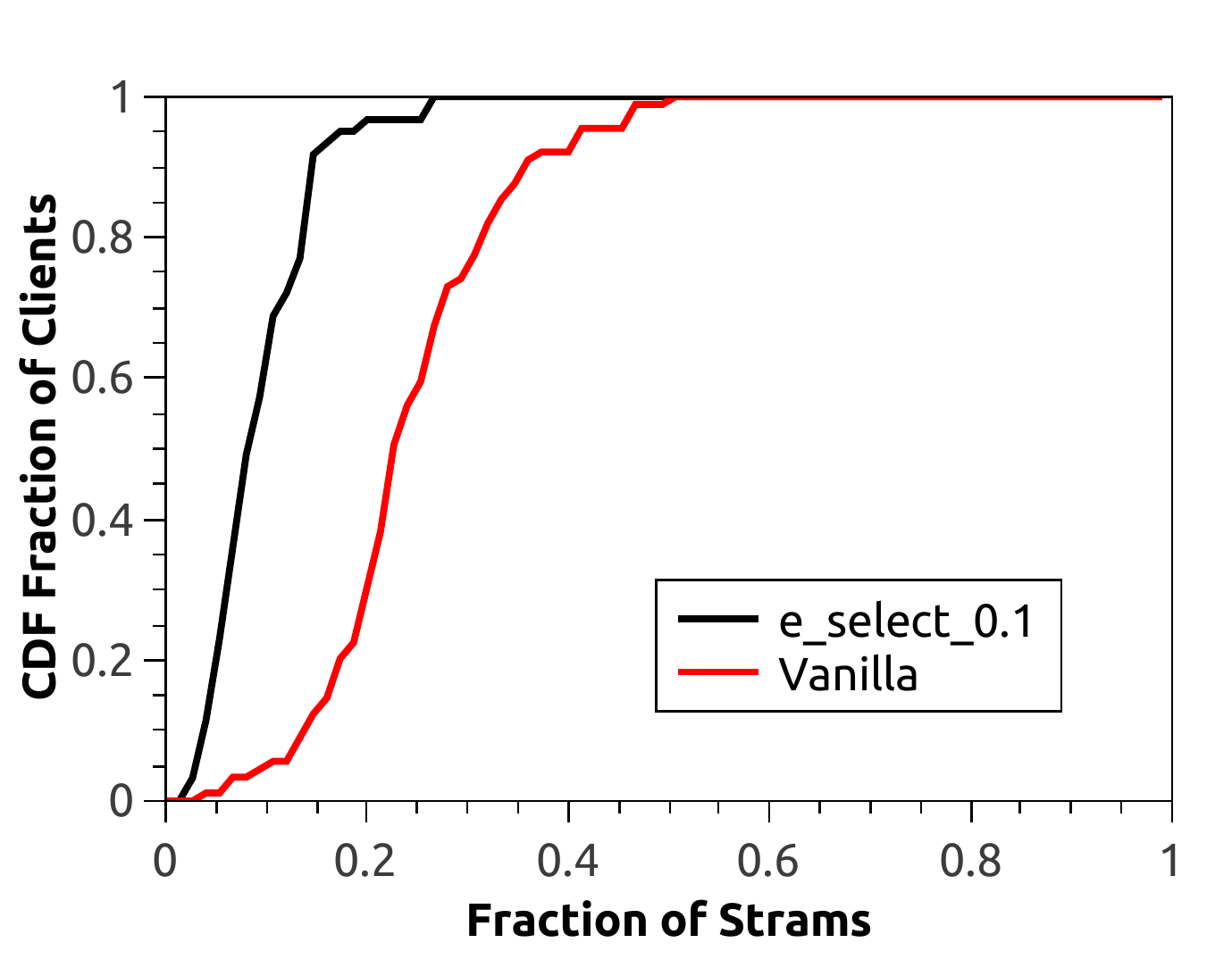}
    \vskip -0.2cm	
    \caption{Fraction of vulnerable streams for DeNASA e-select compared to Vanilla with respect to AS adversaries.}
\label{comprate}
\vspace{-7mm}
\end{figure}

The two methods used in DeNASA are: 1) \textit{e-select}, and 2) \textit{g-select}. \textit{E-select} determines how clients select exit relays based on a tunable parameter \(\tau\) ranging from \(0\) to \(1\).  When \(\tau = 0.1\) clients are restricted to selecting from a smaller set of exits that have lower probability of traversing the suspect ASes. Additionally, the set of exits available for each client is dependent on the client's location. As \(\tau\) increases, the restriction is relaxed.  

Using the described TorPS configuration, we ran a nine month simulation for e-select \(\tau = 0.1\) and Vanilla. We denoted streams as being \textit{vulnerable} if AS3356 or AS1299 appeared on both sides of the stream.  As shown in Fig.~\ref{comprate}, the median vulnerable stream rate for e-select and Vanilla was 8\% and 22\% respectively -- indicating that e-select builds 63\% fewer vulnerable streams compared to Vanilla. 

The \textit{g-select} method ensures that clients only select guards for which there are no suspect ASes in the AS-level path from client to guard. The suspect AS list is tunable such that the client can avoid from one to eight suspect ASes. By avoiding more suspect ASes from the client side, clients maintain a more restrictive set of possible guards to choose. Additionally, the set of guards available for each client is dependent on the client's location.    

\begin{figure}
    \centering
    \vskip -0.5cm	
    \includegraphics[width=3.0in, height=1.7in]{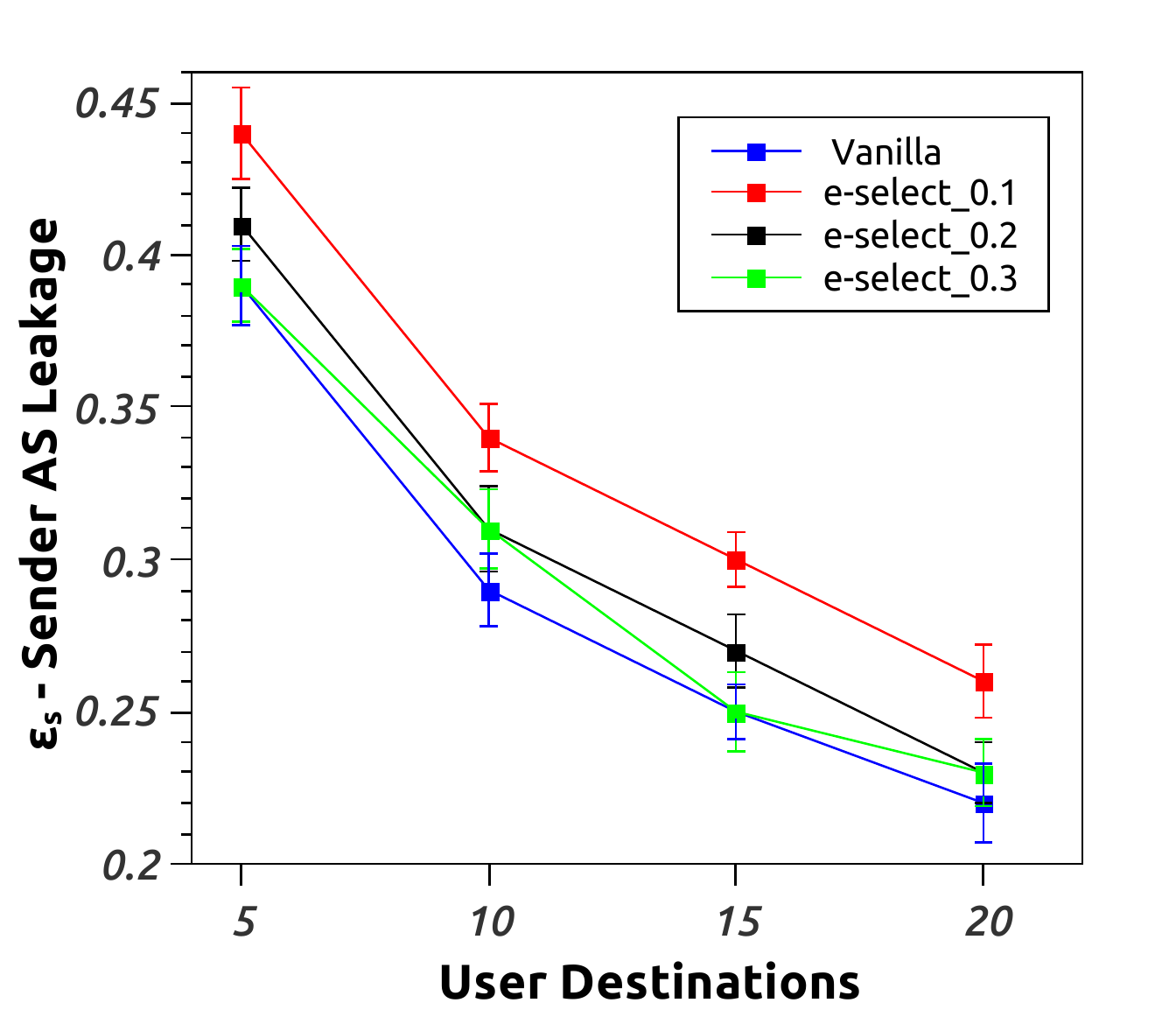}
 %   \vskip -0.2cm	
    \caption{Sender AS leakage for DeNASA exit selection variations compared to Vanilla Tor as a function of user model.}
\label{leakage}
\vspace{-5mm}
\end{figure}
    
\subsection{Experiments}
For each experiment, we generated 1.8 million Tor paths to train the CLASI adversary classification model. We then ran the CLASI challenge-response game for 3,000 new paths on which the adversary made prediction attempts. For each data point, this process was repeated 30 times and we plot the mean along with a 95\% confidence interval. In Figure~\ref{leakage}, we plot sender AS leakage for different variants of \textit{e-select} compared to vanilla Tor. The Figure indicates that sender AS leakage is higher for \textit{e-select} compared to vanilla Tor. Moreover, sender AS leakage increases for \textit{e-select} as the threshold \(\tau\) is decreased.  For example, for the \textit{5-, 10-, 15-}, and \textit{20-destination} user models, sender AS leakage increased by 7\%, 10\%, 11\%, and 13\%, respectively, for \(\tau = 0.1\) compared to \(\tau = 0.2\). We found similar results when \(\tau\) was increased from \(0.2\) to \(0.3\). This was expected due to the set of exits being more restrictive for clients when using lower values for \(\tau\). 
%As expected, decreasing the threshold $\tau$ increases the amount of leakage, since the client is more restricted in the choice of exit nodes. 
%\todo{include some quantitative results. Armon: Done.}

Also as expected, as user destinations increased, sender leakage decreased for all path selection algorithms. More specifically, for Vanilla Tor, sender AS leakage decreased by \(26\%\) for the \textit{10-destination} user model compared to the \textit{5-destination} user model, by \(14\%\) for the \textit{15-destination} user model compared to the \textit{10-destination} user model, and by \(12\%\) for the \textit{20-destination} user model compared to the \textit{15-destination} user model.%\todo{include quantitative results. Armon: Done.}
This highlights the impact that the user model may have on security results.  %We recommend that researchers model several reasonable user models when presenting security evaluation results.   
Moreover, we note that CLASI is sensitive to changing user models and should be a useful tool for researchers seeking to gain more understanding of security implications for their proposed path selection algorithms under different user models.

\begin{figure}
    \centering
%    \vskip -0.6cm	
    \includegraphics[width=3.0in, height=1.6in]{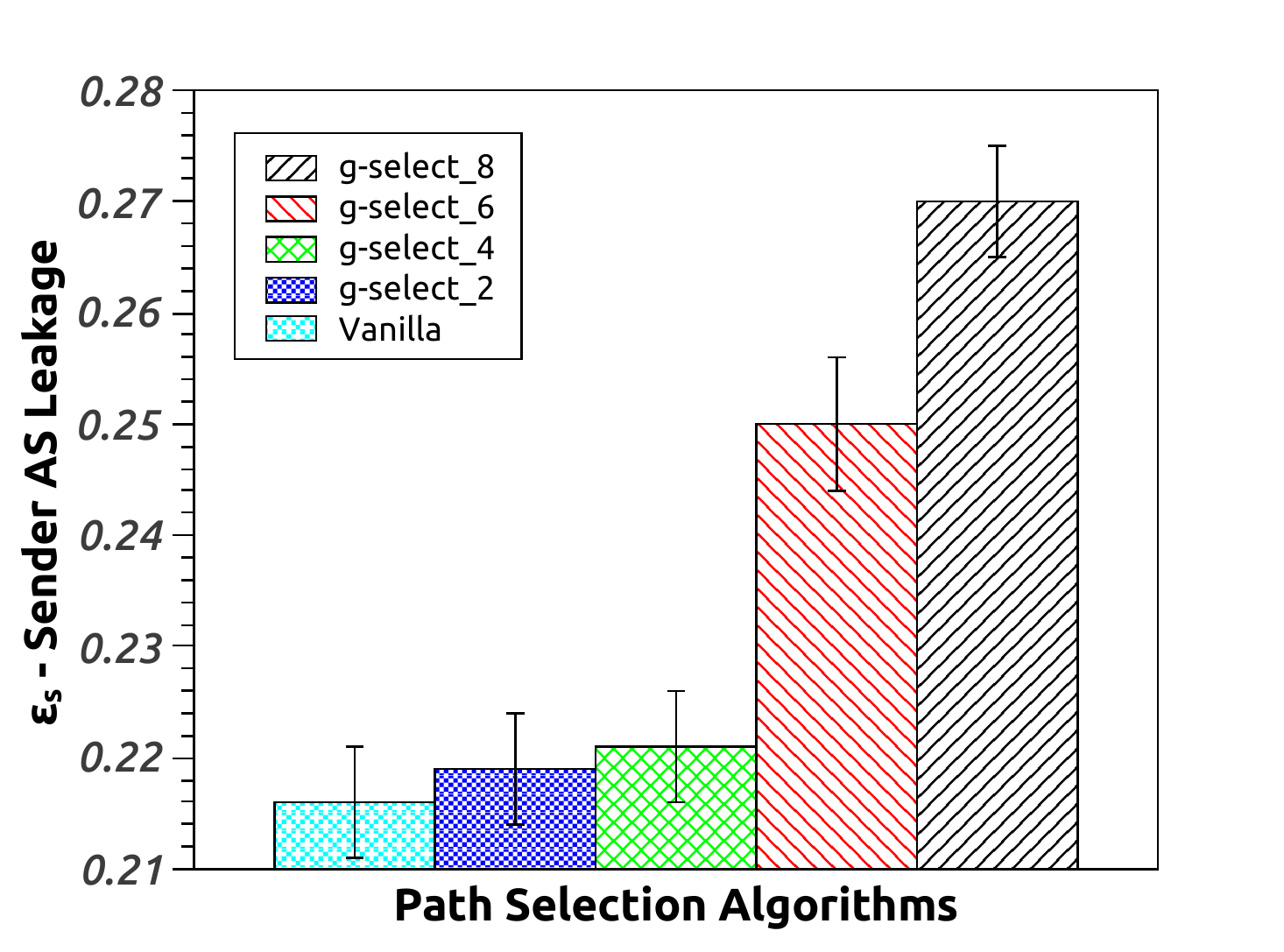}
%    \vskip -0.2cm	
    \caption{Sender AS leakage for DeNASA guard selection variations compared to Vanilla Tor.}
\label{gleakage}
\end{figure}

Figure~\ref{gleakage} shows sender AS leakage for DeNASA guard selection variations compared to Vanilla Tor. Sender AS leakage was not significantly different for Vanilla Tor compared to \textit{g-select} when two to four suspect ASes were avoided. On the other hand, there was a \(14\%\) increase in leakage for when six suspect ASes were avoided compared to when four suspect ASes were avoided. Similarly, there was an \(8\%\) increase in leakage for when eight suspect ASes were avoided compared to when six suspect ASes were avoided.%\todo{quantitative. Armon: Done}

\begin{figure}
    \centering
    %\vskip -0.6cm	
    \includegraphics[width=0.46\textwidth,height=2.0in]{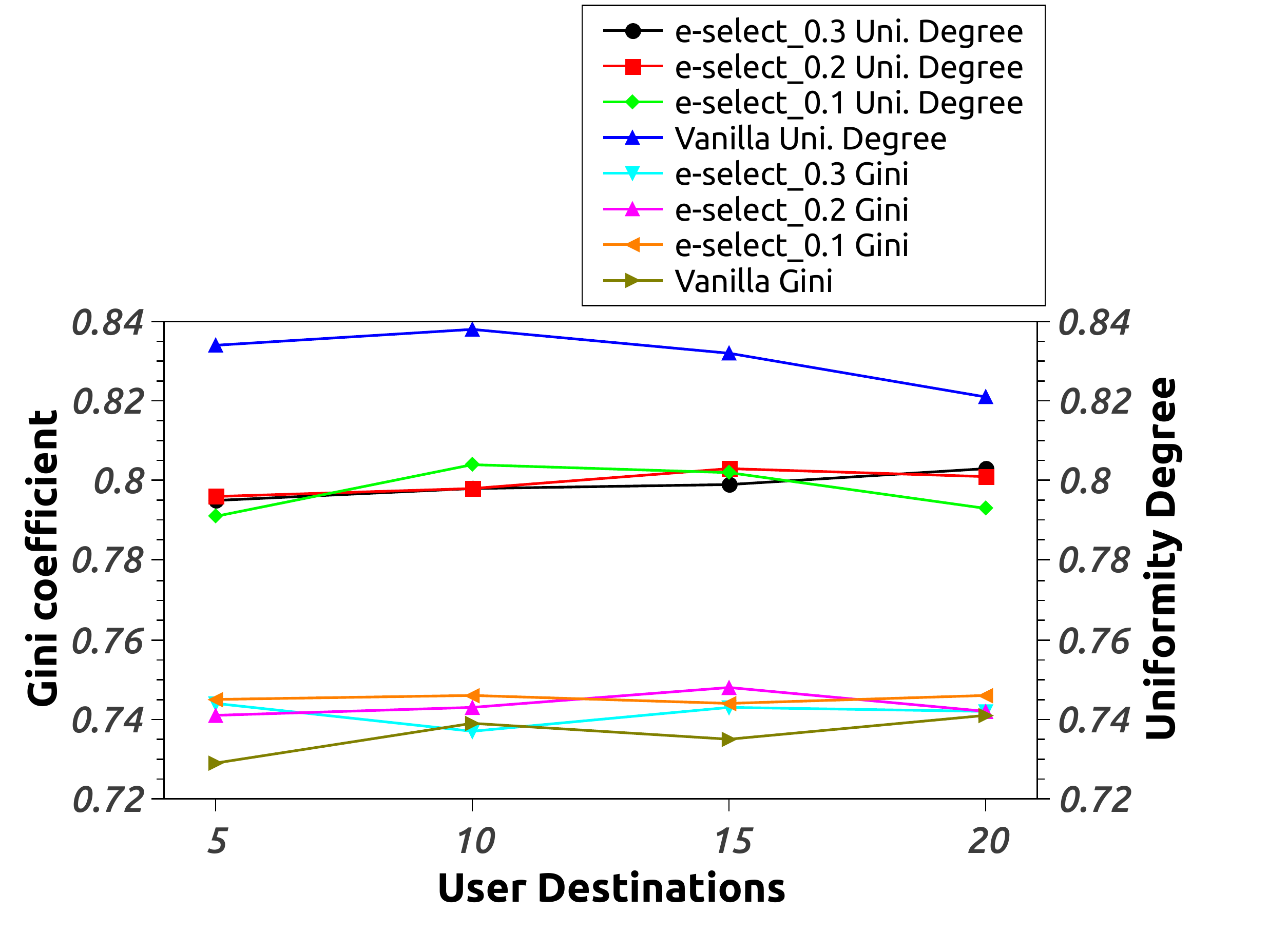}
 %   \vskip -0.2cm	
    \caption{Gini Coefficient and Uniformity degree for DeNASA exit selection variations compared to Vanilla Tor.}
\label{metrics}
\vspace{-5mm}
\end{figure}

\subsection{Entropy Based Metrics}

To understand the value of CLASI as an anonymity metric, it is necessary to see the results of other anonymity metrics. In this section, we show results for DeNASA using two anonymity metrics, Gini coefficient~\cite{snader2008tune} and Uniformity degree~\cite{diaz2003towards}.
%\todo{point to papers that use these as evidence that they are important. Armon: Done.}

In Figure~\ref{metrics} we plot Gini coefficient and uniformity degree for DeNASA exit selection variants compared to Vanilla Tor. The x-axis shows the number of user destinations. The two measures have an inverse relationship. As Gini coefficient grows, anonymity goes down, while as uniformity degree grows, anonymity goes up. We see that both measures show little difference for different values of the threshold $\tau$ or the number of user destinations. In contrast, CLASI does show a significant differences in sender location leakage as these parameters vary. This highlights an advantage for CLASI, in that it can be used by researchers to understand the anonymity impact of path selection algorithms under various user models in Tor.

%As Gini Coefficient and Uniformity Degree increase and decrease respectively, anonymity decreases.  First, since Gini Coefficient and Uniformity Degree are designed to measure anonymity at the relay selection level, they did not show a significant change in anonymity when the user model was varied.  On the other hand, CLASI did show a significant change in anonymity with respect to changing user models.  This result highlights an advantage for CLASI in that it can be used by researchers to understand the anonymity impact of path selection algorithms under various user models in Tor. 

Additionally, there was not a significant change in Gini coefficient for Vanilla compared to DeNASA's exit selection variants. This anomaly highlights a significant disadvantage for Gini coefficient in measuring anonymity for path selection algorithms in Tor. The result is due to the fact that Gini coefficient is a measure of equality of relay selection for all clients in the anonymous communication system taken together. As an extreme example, suppose that all Tor users are split evenly into users from country A and those from country B. Also suppose that all Tor relays are also split into two groups with equal bandwidths. If all users from country A select only relays from the first group and country B users from the second group, then the Gini coefficient will be the same as Vanilla Tor, even though the choice of relays clearly indicates which country the user is in. Thus, Gini coefficient is not suitable for understanding anonymity loss when clients use some bias relevant to their location to select paths.

%As an extreme example, suppose that we split the Tor relays in half, with each half containing half of the available bandwidth for each of three positions on the path (guard, middle, and exit). Then suppose that all the users in Europe select only relays from one half while all users outside of Europe select only relays from the other half, pretending that Europe represents half of the users of Tor.
%For example, if one set of clients always choose one set of relays and another set of clients always choose another set of relays with equal probability, the Gini coefficient will not indicate a loss of anonymity because both sets of relays are chosen with equal frequency.  However, the clients become distinguishable due to their unique relay selection policy.     
There was a significant decrease in uniformity degree for DeNASA's exit selection variants compared to Vanilla Tor. However, there was no significant change in Uniformity Degree with respect to changing values of \(\tau\) for the three DeNASA exit selection variants themselves. On the other hand, CLASI did show a significant difference in anonymity among the three DeNASA exit selection variants. This shows a disadvantage for uniformity degree in measuring anonymity for path selection algorithms in Tor.

\begin{figure}
    \centering
    \vskip -0.6cm	
    \includegraphics[width=3.15in, height=1.8in]{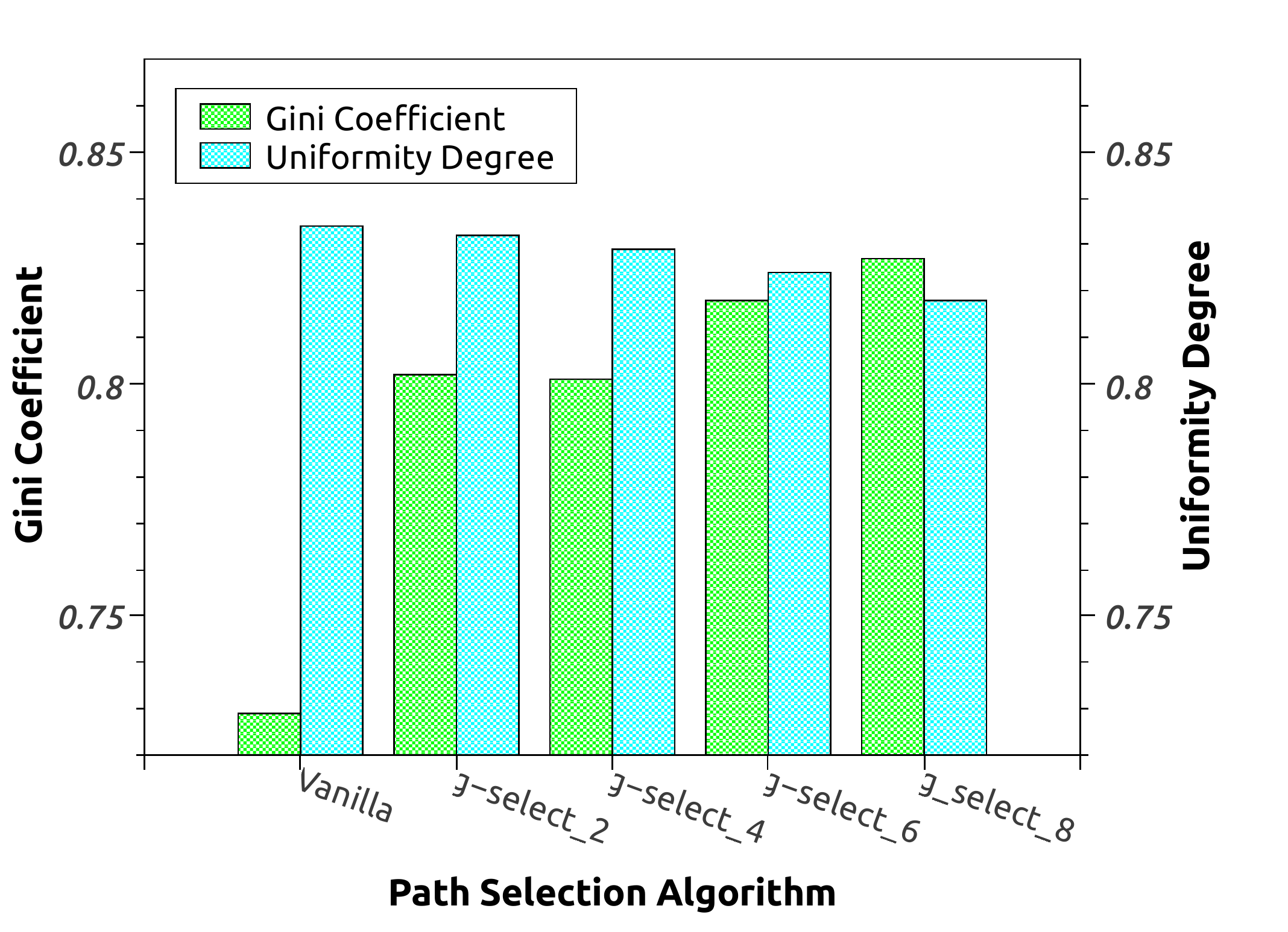}
    \vskip -0.2cm	
    \caption{Gini Coefficient and Uniformity Degree for DeNASA guard selection variants compared to Vanilla Tor.}
\label{gmetrics}
\vspace{-5mm}
\end{figure}

In Figure~\ref{gmetrics} we plot Gini Coefficient and Uniformity Degree for DeNASA's guard selection variants compared to Vanilla Tor for the \textit{5-destination} user model. The results indicate a loss in anonymity, as Gini Coefficient significantly increased for all three g-select variants compared to Vanilla Tor. However, there was no significant change in Gini coefficient among the three variants of g-select.  

There was no significant change in uniformity degree for Vanilla compared to DeNASA's guard selection variants. This shows that uniformity degree did not indicate that there was a \textit{guard placement attack} vulnerability even if clients were configured to avoid up to eight suspect ASes while selecting their guard nodes.

We conclude that gini coefficient and uniformity degree are not sufficient replacements for the CLASI metric when measuring anonymity of path selection algorithms in Tor.

\subsection{Time To First Compromise}
\begin{figure}
    \centering
    \vskip -0.6cm	
    \includegraphics[width=2.8in, height=1.6in]{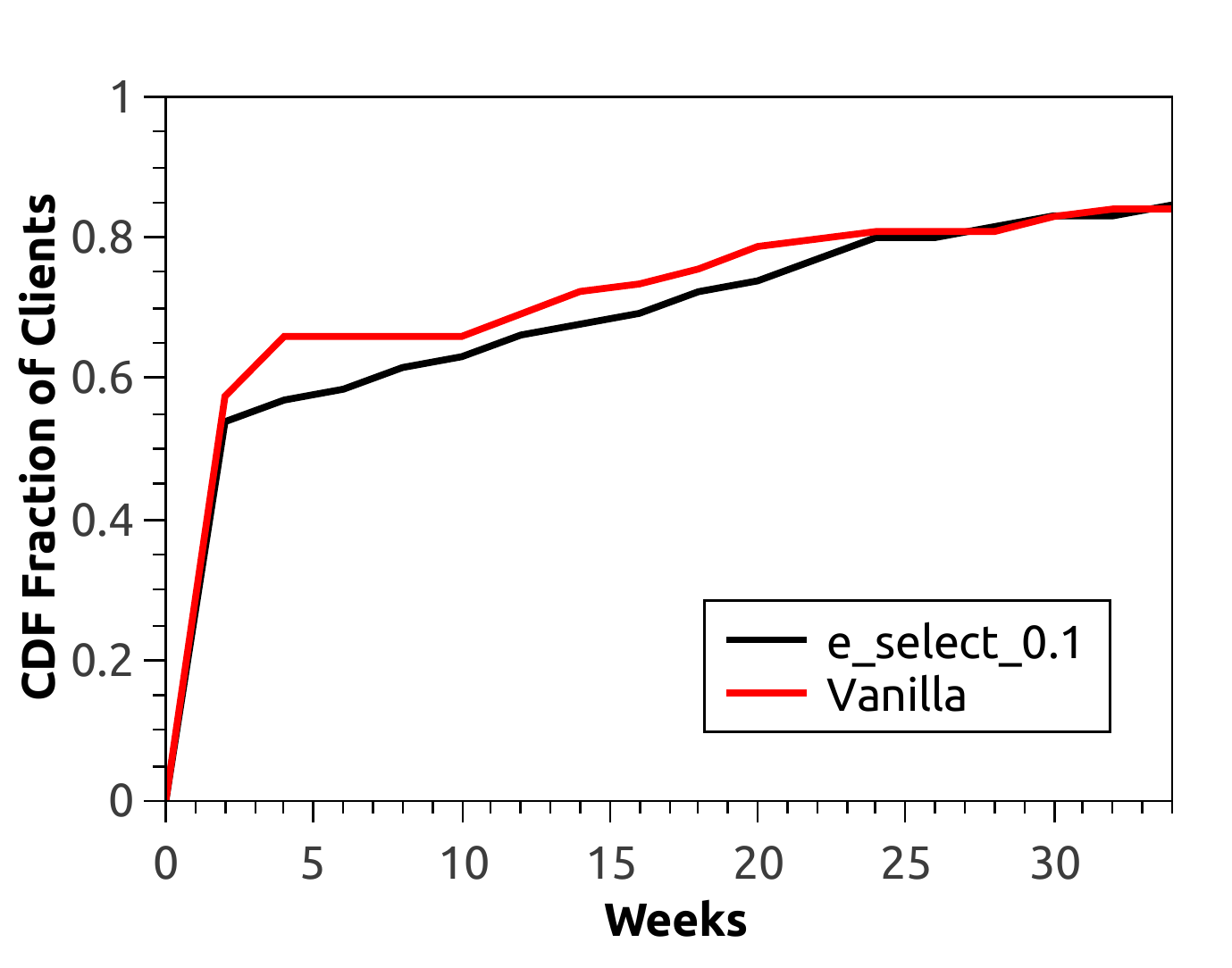}
   \vskip -0.2cm	
    \caption{Time to first compromise for DeNASA e-select compared to Vanilla Tor}
\label{TTFC}
\vspace{-8mm}
\end{figure}

Time to first compromise is an all-or-nothing measure of how long it takes until a client uses a compromised circuit~\cite{johnson2013users}. Using the TorPS configuration described in Section~\ref{torps}, we ran a nine-month simulation for e-select \(\tau = 0.1\) and Vanilla. We denoted streams as being \textit{vulnerable} if AS3356 or AS1299 appeared on both sides of the stream. As shown in Figure~\ref{TTFC}, approximately 60\% of Vanilla and e-select clients built at least one vulnerable stream within the first two weeks. After the nine month period, approximately 80\% of Vanilla and e-select clients built at least one vulnerable stream. On the other hand, according to Figure~\ref{comprate}, DeNASA builds 63\% fewer vulnerable streams with respect to AS adversaries compared to Vanilla. Therefore, some DeNASA clients should realize some security improvement compared to Vanilla clients because they build less vulnerable streams, even though the time taken for DeNASA clients to build their first vulnerable stream is similar to Vanilla Tor. 

The results indicate that DeNASA e-select \(\tau = 0.1\) provides approximately the \textit{same} security against AS-level adversaries with respect to \textit{time to first compromise}, and better security against AS-level adversaries with respect to \textit{vulnerable stream rate}.  Conversely, in Figure~\ref{leakage}, CLASI shows a security \textit{reduction} in client AS leakage of 11\% for e-select \(\tau = 0.1\) compared to Vanilla. These contrasting results support our assertion that \textit{time to first compromise} alone is not sufficient in fully understanding the security implications of path selection algorithms. Though \textit{time to first compromise} is an important metric, we point out that other metrics including CLASI should be used when measuring anonymity for path selection algorithms in Tor.  

%% file: MyFiles/security.tex
\section{PredicTor Security Evaluation}

To understand the anonymity level of PredicTor and PredicTor+CAR compared to CAR and Vanilla, we first generated 500,000 paths for each algorithm from the Shadow experiment described in Section~\ref{shadowsim}. Then, we measured anonymity of each path selection algorithm using Gini coefficent, Uniformity degree, and CLASI.   

\begin{figure}
    \centering
 %   \vskip -0.6cm	
    \includegraphics[width=3.0in, height=1.7in]{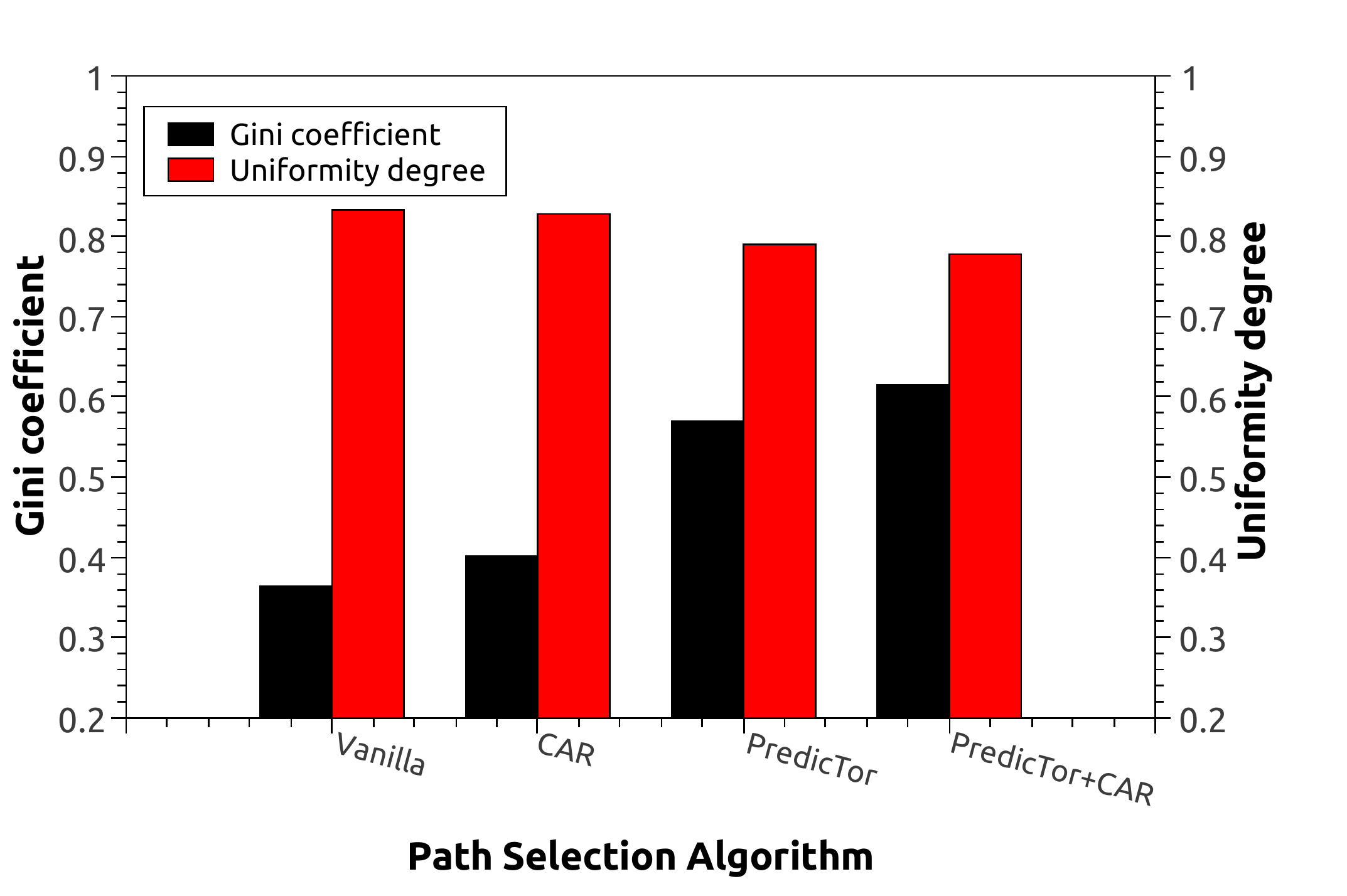}
    \vskip -0.2cm	
    \caption{Gini coefficient and Uniformity degree.}
\label{securityEval_metrics}
\vspace{-5mm}
\end{figure}

Figure~\ref{securityEval_metrics} shows Gini coefficient and Uniformity degree for all tested algorithms.  %There was a slight increase in Gini coefficient for CAR compared to Vanilla. 
The Gini coefficient was 0.21 higher for PredicTor over Vanilla and 0.25 higher for PredicTor+CAR over Vanilla. According to the Gini coefficient metric, Vanilla and CAR had similar anonymity while PredicTor and PredicTor+CAR had significantly worse anonymity. These results suggest that PredicTor clients select some relays with higher probability and avoid other relays, causing an inequality in relay selection compared to Vanilla. In contrast, there was a slight decrease in Uniformity degree for PredicTor and PredicTor+CAR compared to CAR and Vanilla. 
%On the other hand, there was no significant change between PredicTor and PredicTor+CAR, or between Vanilla and CAR.

%There was no significant change in Uniformity degree for Vanilla, CAR, and PredicTor.  These results conflict with Gini coefficient and lead us to infer that, though there is an inequality in relay selection for PredicTor, the overall distribution of selected relays for PredicTor clients is similar to Vanilla Tor.  

%However, Uniformity degree for PredicTor+CAR significantly decreased by \(0.06\).     

%According to Uniformity degree, anonymity among Vanilla, CAR, and PredicTor was similar while anonymity for PredicTor+CAR was significantly worse.

In Figure~\ref{PredicTor_CLASI}, we plot sender AS leakage using CLASI for PredicTor and PredicTor+CAR compared to CAR and Vanilla. We found %$\epsilon_s=0.36 \pm .015$ for Vanilla and \(\epsilon_s=0.38 \pm .016\)  for PredicTor. We infer that 
that PredicTor clients had similar AS leakage compared to Vanilla, likely due to clients choosing paths independently of their location in PredicTor.  %path selection for PredicTor is homogeneous for all clients.  This characteristic allows PredicTor clients to be concealed within their anonymity set. 
On the other hand, CAR clients build paths based on opportunistic measurement from the client, and thus, their paths should have some relationship to their location. Accordingly, we found a slight increase in AS leakage for CAR and a significant increase of about 9\% for PredicTor+CAR compared to Vanilla and PredicTor alone. We note that both Gini coefficient and Uniformity degree did not indicate a significant difference in anonymity for PredicTor+CAR compared to PredicTor. 
%On the other hand, we measured an increase of \(8.8\%\) in client AS leakage for PredicTor+CAR compared to PredicTor.

%We measured an \(\epsilon_s\) value of \(0.38 \pm .015\) and \(0.38 \pm .016\) for Vanilla and PredicTor respectively., while \(\epsilon_s\) for Vanilla was \(0.36 \pm .026\).  According to our findings, there was a significant increase in sender AS leakage for CAR and PredicTor compared to Vanilla, however CAR and PredicTor both had similar AS leakage.

From our findings, we conclude that AS leakage for PredicTor is similar to Vanilla and slightly better than CAR due to PredicTor clients building paths independently of network location. On the other hand, PredicTor does select certain relays with higher probability causing an inequality in relay selection compared to Vanilla and CAR. However, the entropy loss is minimal, indicating that the distribution of selected relays for all PredicTor clients is similar to Vanilla.

\paragraphX{Time to First Compromise}  Johnson et al.~\cite{johnson2013users} showd that time to first compromise is strongly related to guard selection policy.  As PredicTor chooses guards exactly the same as Vanilla, we believe time to first compromise for PredicTor due to relay-level and AS-level adversaries should be similar to Vanilla Tor.

%% file: MyFiles/disc.tex
\section{Discussion and Future Work}~\label{disc}
In this section, we discuss deployment ideas for PredicTor and future work possibilities for CLASI. 

\subsection{PredicTor Deployment}\label{fwork}
There are two main challenges that would need to be addressed for the successful deployment of PredicTor: 1) clients should routinely receive  comprehensive training data, and 2) the training data should be gathered securely, such that an adversary has little chance of directing traffic to malicious relays. 

In the Live Tor experiments, we built a training set by measuring download times for approximately 50,000 streams from a centralized authority over the course of one hour. This training set was given to a PredicTor client and used to build circuits during the subsequent hour. Gathering the training data added an additional load on the Network of approximately 4.0 GiB/hr, or just .07 GiB/s. The current bandwidth of the Tor Network is approximately 60 GiB/s. Thus, we believe that the measurement load should not be a problem for deployment.

Similar to the Tor consensus, the training set should be sent to each client once per hour. We believe this should not be problematic because our training set was approximately 233 KB. Adding this information to the consensus would result in a file size increase of only 10\%. 

In a live deployment of PredicTor, the training data can be gathered by one single authority or by multiple authorities. If the data is gathered by one authority, then that authority should be trusted. If the training data is gathered by multiple authorities, then there should be a voting process that is used to resist manipulation from a subset of malicious authorities.

Additionally, PredicTor measurement circuits should be made indistinguishable from regular circuits by randomizing the destination domain and payload size of the measured stream. We note that using RTT measurements for performance is fundamentally insecure because they can easily be manipulated by malicious exit relays.

%\paragraphX{Simulation Results}
%Our PredicTor results were acquired by aggregating training data from a Tor network simulation using Shadow. Then, we tested PredicTor in a different simulation instance using the same Tor network configuration. One idea for future work would be to test PredicTor on the live Tor network in order to add even more validity to our results. On the other hand, we believe our results are valid because they support the findings of Wang et al.~\cite{wang2012congestion} that some Tor relays experience persistent congestion. Moreover, better performance can be achieved by avoiding persistently congested nodes and choosing persistently non-congested nodes. 

\begin{figure}
    \centering
%    \vskip -0.6cm	
    \includegraphics[width=3.0in, height=1.7in]{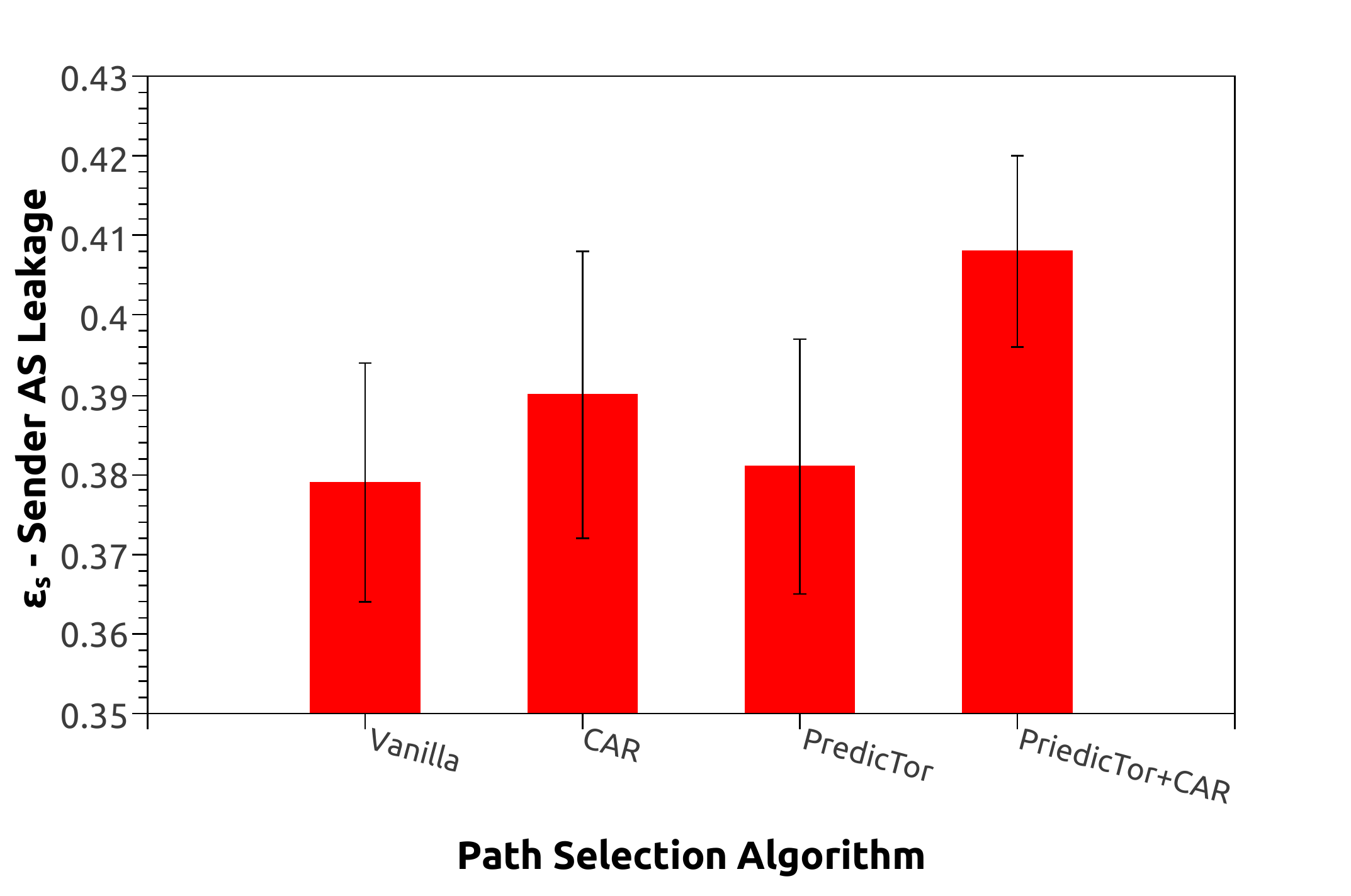}
   \vskip -0.2cm	
    \caption{Sender AS leakage}
\label{PredicTor_CLASI}
\vspace{-5mm}
\end{figure}

%\paragraphX{Deployment.}
%In PredicTor, comprehensive training data is the critical step for clients to have global congestion knowledge about the Tor network. As such, one critical step to achieve the results that we have presented in this paper is to ensure PredicTor clients receive comprehensive training data. 
%More ideas about providing training data to clients are discussed in this section. In the deployment of PredicTor, t
%There are two methods in which clients can acquire training data.  In the first method, the \textit{Directory Authorities} can run Tor network Shadow simulations similar to the simulation we presented in this paper. Then, the authorities would send the training data to each client in the consensus. A disadvantage of this approach is that it would result in a larger consensus document. 

%Another option is that clients can opportunistically build their own training sets over time by aggregating historical circuit download times. A disadvantage to this is that clients would need to be connected for some time before experiencing better performance. Additionally, this method may decrease anonymity, as some clients may build circuits based on their location.

%Perhaps a combination of the two stated methods could be used to mitigate some disadvantages of both. We leave these specifics open for future work.
\subsection{CLASI}
\paragraphX{Adversary Model.} The adversary model within the CLASI challenge-response game is an all-knowing adversary. Therefore, our results yield an upper bound. Meiser et al.~\cite{backes2016your} showed that a budget adversary model results in a tight upper bound. The CLASI adversary model can be modified such that the adversary's knowledge is bounded by their budget. The budget can be defined in terms of cost or bandwidth, for example. 

\paragraphX{Sender/Receiver Anonymity.} CLASI is designed to measure AS leakage from the sender.  However, the classification model can be modified to also measure AS leakage from the receiver. This could give researchers even more insight into anonymity implications, especially for \textit{destination-aware} path selection algorithms that use destination information to build circuits~\cite{starov2015measuring,edman2009awareness}.  
\vspace{-2mm}

%% file: MyFiles/conclusion.tex
\section{Conclusion}
To address Tor performance, we presented {\em PredicTor}, a path selection technique that uses a Random Forest classifier trained on a set of recent Tor paths to predict the performance of a proposed path. We implemented PredicTor in the Tor source code and showed through simulations in Shadow that PredicTor improved Tor network performance by 23\% compared to Vanilla Tor and by 13\% compared to Congestion-Aware Routing. In our live Tor experiments, during times of high congestion, PredicTor had an improvement of 7\% to 13\% in the median case compared to Vanilla Tor. We evaluated the anonymity of PredicTor using standard entropy-based metrics, and we proposed a new anonymity metric called \textit{CLASI}: Client Autonomous System Inference. Our results indicated that CLASI showed anonymity loss for location-aware path selection algorithms where other entropy based metrics showed little to no loss of anonymity. Additionally, CLASI indicated that PredicTor had similar client AS leakage compared to Vanilla due to PredicTor building circuits that are independent of client location.

%% file: MyFiles/appendix.tex
%\pagebreak
\clearpage
\section*{Appendices}

\section{Client Location and Guard Diversity}~\label{a1}

In Table~\ref{tab:1}, we show the median and 90th percentile performance and circuit distance improvements for PredicTor compared to Vanilla.

\begin{table}[h]
\renewcommand{\arraystretch}{1.3}
\centering
\begin{tabular}{  c  c  c  c  c  c  c  c  }

  \multirow{2}{*}{Date \& Time} & \multirow{2}{*}{CC} & \multirow{ 2}{*}{Guard} & \multirow{ 2}{*}{Cong.} & \multicolumn{2}{c}{Time} & \multicolumn{2}{c}{Distance} \\ 
   &  &  &  & Median & 90th per. & Median & 90th per. \\ \hline
  2017-07-18 00:00 & US & Fast & Low & $4.2\%$ & $15.6\%$ & $52.0\%$ & $16.5\%$ \\ 
  2017-07-19 14:00 & US & Fast & High & $9.7\%$ & $17.7\%$ & $23.0\%$ & $10.8\%$ \\ 
  2017-09-01 00:00 & US & Slow & Low & $4.4\%$ & $6.3\%$ & $2.2\%$ & $-1.2\%$ \\ 
  2017-08-31 14:00 & US & Slow & High & $6.7\%$ & $19.1\%$ & $-1.0\%$ & $-5.2\%$ \\ \hline
  
  2017-08-15 00:00 & DE & Fast & Low & $7.3\%$ & $23.1\%$ & $29.4\%$ & $16.2\%$ \\ 
  2017-08-03 14:00 & DE & Fast & High & $12.8\%$ & $25.3\%$ & $26.2\%$ & $21.8\%$ \\ 
  2017-08-22 00:00 & DE & Slow & Low & $3.3\%$ & $2.6\%$ & $2.0\%$ & $5.9\%$ \\
  2017-08-17 14:00 & DE & Slow & High & $6.9\%$ & $16.9\%$ & $0.0\%$ & $0.0\%$ \\ \hline
  
  2017-08-30 00:00 & JP & Fast & Low & $7.4\%$ & $11.2\%$ & $28.8\%$ & $18.0\%$ \\ 
  2017-08-29 14:00 & JP & Fast & High & $6.3\%$ & $10.8\%$ & $11.3\%$ & $10.9\%$ \\ 
  2017-08-24 00:00 & JP & Slow & Low & $7.2\%$ & $4.5\%$ & $1.8\%$ & $0.0\%$ \\ 
  2017-08-25 14:00 & JP & Slow & High & $7.3\%$ & $14.0\%$ & $2.1\%$ & $4.1\%$ \\ \hline

\end{tabular}
\vskip -0.2cm
\caption{Improvements in download time and circuit distance for PredicTor compared to Vanilla.}
\label{tab:1}
\vspace{-5mm}
\end{table}

%% file: main.bbl
\begin{thebibliography}{10}

\bibitem{akhoondi2012lastor}
{\sc Akhoondi, M., Yu, C., and Madhyastha, H.~V.}
\newblock {LASTor: A} low-latency {AS}-aware {Tor} client.
\newblock In {\em Proceedings of the 33rd IEEE Symposium on Security and
  Privacy (S\&P'12)\/} (2012).

\bibitem{topalexa}
{\sc Alexa.com.}
\newblock Alexa top sites., 2017.
\newblock http://www.alexa.com/topsites.

\bibitem{annessi2016navigator}
{\sc Annessi, R., and Schmiedecker, M.}
\newblock {Navigator: Finding Faster Paths to Anonymity}.
\newblock In {\em IEEE European Symposium on Security and Privacy (EuroS\&P'
  16)\/} (2016).

\bibitem{backes2013anoa}
{\sc Backes, M., Kate, A., Manoharan, P., Meiser, S., and Mohammadi, E.}
\newblock {AnoA}: {A Framework For Fnalyzing Anonymous Communication
  Protocols}.
\newblock In {\em Proceedings of the 26th IEEE Computer Security Foundations
  Symposium Computer (CSF'13)\/} (2013).

\bibitem{backes2016your}
{\sc Backes, M., Meiser, S., and Slowik, M.}
\newblock {Your Choice MATor (s)}.
\newblock In {\em Proceedings on Privacy Enhancing Technologies (PETS'16)}.

\bibitem{barton2016denasa}
{\sc Barton, A., and Wright, M.}
\newblock Denasa: Destination-naive {AS}-{A}wareness in {A}nonymous
  {C}ommunications.
\newblock In {\em Proceedings on Privacy Enhancing Technologies (PETS'16)}.

\bibitem{borisov2007denial}
{\sc Borisov, N., Danezis, G., Mittal, P., and Tabriz, P.}
\newblock {Denial of Service or Denial of Security?}
\newblock In {\em Proceedings of the 14th ACM conference on Computer and
  communications security (CCS'07)\/} (2007).

\bibitem{breiman2001random}
{\sc Breiman, L.}
\newblock {Random Forests}.
\newblock vol.~Vol 45, Springer.

\bibitem{caidaASrank}
{\sc CAIDA}.
\newblock Caida as ranking.
\newblock http://as-rank.caida.org/.

\bibitem{crovella1998heavy}
{\sc Crovella, M.~E., Taqqu, M.~S., and Bestavros, A.}
\newblock Heavy-tailed probability distributions in the world wide web.
\newblock {\em A practical guide to heavy tails 1\/} (1998), 3--26.

\bibitem{diaz2003towards}
{\sc Diaz, C., Seys, S., Claessens, J., and Preneel, B.}
\newblock {Towards Measuring Anonymity}.
\newblock In {\em Proceedings of the 2nd International Conference on Privacy
  Enhancing Technologies (PET'02)\/} (2002).

\bibitem{dingledine2014one}
{\sc Dingledine, R., Hopper, N., Kadianakis, G., and Mathewson, N.}
\newblock {One Fast Guard For Life (or 9 Months)}.
\newblock In {\em Proceedings of 7th Workshop on Hot Topics in Privacy
  Enhancing Technologies (HotPETs'14)\/} (2014).

\bibitem{dingledine2004tor}
{\sc Dingledine, R., Mathewson, N., and Syverson, P.}
\newblock {Tor: The} {Second-Generation Onion Router}.
\newblock In {\em Proceedings of the 13th USENIX Security Symposium (USENIX
  Security'04)\/} (2004).

\bibitem{edman2009awareness}
{\sc Edman, M., and Syverson, P.}
\newblock {{AS}-Awareness in {Tor} Path Selection}.
\newblock In {\em Proceedings of the 16th ACM conference on Computer and
  communications security (CCS'09)\/} (2009).

\bibitem{geddes2013low}
{\sc Geddes, J., Jansen, R., and Hopper, N.}
\newblock {How Low Can You Go: Balancing Performance with Anonymity in Tor}.
\newblock In {\em International Symposium on Privacy Enhancing Technologies
  Symposium (PETS' 13)\/} (2013).

\bibitem{geddes2016abra}
{\sc Geddes, J., Schliep, M., and Hopper, N.}
\newblock {ABRA CADABRA}: {Magically Increasing Network Utilization in Tor by
  Avoiding Bottlenecks}.
\newblock In {\em Proceedings of the 2016 ACM on Workshop on Privacy in the
  Electronic Society (WPES'16)\/} (2016).

\bibitem{imani2018guardsets}
{\sc Imani, M., Barton, A., and Wright, M.}
\newblock {Guard Sets in Tor using AS Relationships}.
\newblock In {\em International Symposium on Privacy Enhancing Technologies
  Symposium (PETS' 18)\/} (2018).

\bibitem{jansen2012methodically}
{\sc Jansen, R., Bauer, K.~S., Hopper, N., and Dingledine, R.}
\newblock {Methodically Modeling the {Tor} Network}.
\newblock In {\em Proceedings of the 5th USENIX Conference on Cyber Security
  Experimentation and Test (CSET'12)\/} (2012).

\bibitem{jansennever}
{\sc Jansen, R., Geddes, J., Wacek, C., Sherr, M., and Syverson, P.}
\newblock {Never Been {KIST: Tor'}s Congestion Management Blossoms with
  Kernel-Informed Socket Transport}.
\newblock In {\em Proceedings of the 23rd USENIX Security Symposium (USENIX
  Security'14)}.

\bibitem{jansen2011shadow}
{\sc Jansen, R., and Hopper, N.}
\newblock Shadow: {Running Tor} in a {Box For Accurate And Efficient
  Experimentation}.
\newblock In {\em Proceedings of the Network and Distributed System Security
  Symposium (NDSS'12)}.

\bibitem{jansen2016safe}
{\sc Jansen, R., and Johnson, A.}
\newblock {Safely Measuring Tor}.
\newblock In {\em Proceedings of the 23rd ACM SIGSAC Conference on Computer and
  Communications Security (CCS'16)\/} (2016).

\bibitem{jansen2016safely}
{\sc Jansen, R., and Johnson, A.}
\newblock Safely measuring tor.
\newblock In {\em Proceedings of the 2016 ACM SIGSAC Conference on Computer and
  Communications Security\/} (2016), ACM, pp.~1553--1567.

\bibitem{johnson2015avoiding}
{\sc Johnson, A., Jansen, R., Jaggard, A.~D., Feigenbaum, J., and Syverson, P.}
\newblock {Avoiding The Man on the Wire: Improving Tor's Security with
  Trust-Aware Path Selection}.
\newblock In {\em Proceedings of the Network and Distributed System Security
  Symposium (NDSS'17)\/} (2017).

\bibitem{johnson2013users}
{\sc Johnson, A., Wacek, C., Jansen, R., Sherr, M., and Syverson, P.}
\newblock {Users Get Routed: {Traffic} Correlation on {Tor} by Realistic
  Adversaries}.
\newblock In {\em Proceedings of the 20th ACM SIGSAC conference on Computer \&
  communications security (CCS'13)\/} (2013).

\bibitem{juen2012protecting}
{\sc Juen, J.}
\newblock {Protecting Anonymity in the Presence of Autonomous System and
  {Internet} Exchange Level Adversaries}.

\bibitem{TorMetrics}
{\sc Metrics, T.}
\newblock Tor metrics, June 2015.
\newblock https://metrics.torproject.org.

\bibitem{mitchell1997machine}
{\sc Mitchell, T.}
\newblock {\em {Machine Learning}}.
\newblock McGraw-Hill, 1997.

\bibitem{murdoch2008metrics}
{\sc Murdoch, S.~J., and Watson, R.~N.}
\newblock {Metrics for Security and Performance in Low-Latency Anonymity
  Systems}.
\newblock In {\em Proceedings of the 8th International Symposium on Privacy
  Enhancing Technologies (PETS'08)}.

\bibitem{overlier2006locating}
{\sc Overlier, L., and Syverson, P.}
\newblock {Locating Hidden Servers}.
\newblock In {\em Proceedings of the 27th IEEE Symposium on Security and
  Privacy (S\&P'06)}.

\bibitem{rochet2016waterfiling}
{\sc Rochet, F., and Pereira, O.}
\newblock {Waterfiling: Balancing the Tor Network with Maximum Diversity}.
\newblock In {\em Proceedings on Privacy Enhancing Technologies (PETS'17)}.

\bibitem{serjantov2002towards}
{\sc Serjantov, A., and Danezis, G.}
\newblock {Towards an Information Theoretic Metric for Anonymity}.
\newblock In {\em Proceedings of Privacy Enhancing Technologies Workshop
  (2002)}.

\bibitem{shannon2001mathematical}
{\sc Shannon, C.~E.}
\newblock {A Mathematical Theory of Communication}.
\newblock {\em ACM SIGMOBILE Mobile Computing and Communications Review 5}, 1
  (2001), 3--55.

\bibitem{sherr2009scalable}
{\sc Sherr, M., Blaze, M., and Loo, B.~T.}
\newblock {Scalable Link-Based Relay Selection for Anonymous Routing}.
\newblock In {\em Proceedings of the 9th International Symposium on Privacy
  Enhancing Technologies (PETS'09)\/} (2009).

\bibitem{snader2008tune}
{\sc Snader, R., and Borisov, N.}
\newblock {A Tune-up for Tor: Improving Security and Performance in the Tor
  Network}.
\newblock In {\em Proceedings of the 16th Annual Network \& Distributed System
  Security Symposium (NDSS'08)\/} (2008).

\bibitem{starov2015measuring}
{\sc Starov, O., Nithyanand, R., Zair, A., Gill, P., and Schapira, M.}
\newblock Measuring and mitigating {AS-}level adversaries against {Tor}.
\newblock In {\em Proceedings of the 24th Annual Network \& Distributed System
  Security Symposium (NDSS'16)}.

\bibitem{sun2017counter}
{\sc Sun, Y., Edmundson, A., Feamster, N., Chiang, M., and Mittal, P.}
\newblock {Counter-RAPTOR: Safeguarding Tor Against Active Routing Attacks}.
\newblock In {\em Proceedings of the 38th IEEE Symposium on Security and
  Privacy (S\&P'17)\/} (2017).

\bibitem{sun2015raptor}
{\sc Sun, Y., Edmundson, A., Vanbever, L., Li, O., Rexford, J., Chiang, M., and
  Mittal, P.}
\newblock {RAPTOR: Routing} {A}ttacks on {P}rivacy in {Tor}.
\newblock In {\em Proceedings of the 25th USENIX Security Symposium (USENIX
  Security'15)\/} (2015).

\bibitem{syverson2009m}
{\sc Syverson, P.}
\newblock {Why I'm not an entropist}.
\newblock In {\em International Workshop on Security Protocols\/} (2009),
  Springer, pp.~213--230.

\bibitem{torps}
{\sc TorPS.}
\newblock {TorPS: The Tor} path simulator., 2013.
\newblock http://torps.github.io.

\bibitem{vincenty1975direct}
{\sc Vincenty, T.}
\newblock {Direct and Inverse Solutions of Geodesics on the Ellipsoid with
  Application of Nested Equations}.
\newblock In {\em Survey Review\/} (1975).

\bibitem{wacek2013empirical}
{\sc Wacek, C., Tan, H., Bauer, K.~S., and Sherr, M.}
\newblock {An Empirical Evaluation of Relay Selection in {Tor}}.
\newblock In {\em Proceedings of the 20th Annual Network \& Distributed System
  Security Symposium (NDSS'13)\/} (2013).

\bibitem{wang2012congestion}
{\sc Wang, T., Bauer, K., Forero, C., and Goldberg, I.}
\newblock {Congestion-Aware Path Selection for Tor}.
\newblock In {\em Proceedings of the 16th International Conference on Financial
  Cryptography and Data Security (FC'12)\/} (2012).

\bibitem{wright2003defending}
{\sc Wright, M., Adler, M., Levine, B.~N., and Shields, C.}
\newblock {Defending Anonymous Communications Against Passive Logging Attacks}.
\newblock In {\em Proceedings of the 24th IEEE Symposium on Security and
  Privacy (S\&P'03)\/} (2003).

\bibitem{wright2008passive}
{\sc Wright, M.~K., Adler, M., Levine, B.~N., and Shields, C.}
\newblock {Passive-Logging Attacks Against Anonymous Communications Systems}.
\newblock {\em ACM Transactions on Information and System Security (TISSEC)
  11}, 2 (2008), 3.

\end{thebibliography}
